\documentclass[twocolumn,superscriptaddress,letter]{revtex4}%
\usepackage{amssymb}
\usepackage{color}
\usepackage{graphicx}
\usepackage{dcolumn}
\usepackage{bm}
\usepackage[header,title,page,titletoc]{appendix}
\usepackage{amsmath}
\usepackage{amsfonts}%
\providecommand{\U}[1]{\protect \rule{.1in}{.1in}}

\begin{document}
\title{Composite Spin Liquid in Correlated Topological Insulator - Spin Liquid
without Spin-Charge Separation}
\author{Jing He}
\affiliation{Department of Physics, Beijing Normal University, Beijing, 100875 P. R. China }
\author{Ying Liang}
\affiliation{Department of Physics, Beijing Normal University, Beijing, 100875 P. R. China }
\author{Su-Peng Kou}
\thanks{Corresponding author}
\email{spkou@bnu.edu.cn}
\affiliation{Department of Physics, Beijing Normal University, Beijing, 100875 P. R. China }

\begin{abstract}
In this paper, we found a new type of insulator --- \emph{composite spin
liquid} which can be regarded as a short range B-type topological
spin-density-wave proposed in Ref.\cite{he2}. Composite spin liquid is
topological ordered state beyond the classification of traditional spin liquid
states. The elementary excitations are the "composite electrons" with both
spin degree of freedom and charge degree of freedom, together with topological
spin texture. This topological state supports chiral edge mode but no
topological degeneracy.

PACS numbers: 71.10.Pm, 75.10.Kt, 73.43.Cd, 71.27.+a, 05.30.Pr

\end{abstract}
\maketitle

\section{Introduction}

The Fermi liquid based view of the electronic properties has been very
successful as a basis for understanding the physics of conventional solids
including metals and (band) insulators. For the band insulators, due to the
energy gap, the charge degree of freedom is frozen. For magnetic insulators
with spontaneous spin rotation symmetry breaking, the elementary excitations
are the gapped quasi-particle (an electron or a hole) that carry both spin and
charge degree of freedoms and the gapless spin wave (the Goldstone mode). For
this case, the global symmetry is broken from \textrm{SU(2)} down to
\textrm{U(1)}. Thus the low energy effective model is an \textrm{O(3)}
nonlinear $\sigma$-model ($\mathrm{NL}\sigma \mathrm{M}$) that describes long
wave spin fluctuations.

However, in some special insulators with spin-rotation symmetry and
translation symmetry, due to a big energy gap of electrons, the charge degree
of freedom is totally frozen, emergent gauge fields and deconfined spinons
(the elementary excitation with only spin degree of freedom of an electron)
may exist. People call them \emph{quantum spin liquids}\cite{pw}. People have
been looking for quantum spin liquid states in spin models for more than two
decades \cite{Fazekas,wen,wxg}. In particular spin models, the quantum spin
liquids are accessed (in principle) by appropriate frustrating interactions.
In general there exist three types of ansatz of spin liquid: $SU(2)$, $U(1)$
and $Z_{2}$\cite{wen,wxg}. The three different states may have the same global
symmetry, as conflicts to Landau's theory, in which two states with the same
symmetry belong to the same phase. Since one cannot use symmetry and order
parameter to describe quantum orders, a new mathematical object - projective
symmetry group (PSG) - was introduced\cite{wen,wxg} to characterize the
quantum order of spin liquid states.

Recently, people look for spin liquids in the generalized Hubbard model of the
intermediate coupling region, for example, the Hubbard model on the triangular
lattice, the Hubbard model on the honeycomb lattice, the $\pi$-flux Hubbard
model on square lattice\cite{Lee,Hermele,kou1}. And, the quantum spin liquid
state near Mott transition (MI) of the Hubbard model on honeycomb lattice has
been confirmed by different approaches\cite{meng,kou4,z21,z22,z23,z24,z25}.
However, the nature of the spin liquid in the generalized Hubbard model of the
intermediate coupling region is still debated.

In this paper we found that there may exist another type of insulator with
spin-rotation symmetry and translation symmetry, of which the elementary
excitation has both spin degree of freedom and charge degree of freedom. We
call it \emph{composite spin liquid}. Composite spin liquid (SL) can be
regarded as short range B-type topological spin-density-wave (B-TSDW) which is
beyond the classification of traditional spin liquid states. In a composite
SL, there is no spin-charge separation : the elementary excitation is
so-called "\emph{composite electron}" - a spin one-half charge $\pm e$ object
trapping a topological spin texture (skyrmion or anti-skyrmion). In addition,
the composite SL is a topological spin liquid state with chiral edge states.
However, similar to the case of integer quantum Hall state, composite SL has
no topological degeneracy for the ground state.

The paper is organized as follows. Firstly, we write down the Hamiltonian of
the topological Hubbard model. Secondly we derive the effective \textrm{O(3)}
nonlinear $\sigma$ model with the Chern-Simons-Hopf (CSH) term to learn its
properties. Next, chiral SL and composite SL are found to be the ground state
of the short range A-type topological spin-density-wave and short range B-type
topological spin-density-wave, respectively. Finally, the conclusions are
given. In addition we compare composite SL with other exotic quantum states
including fractional quantum Hall states, spin liquids and topological insulators.

\section{Model and mean field results}

The Hamiltonian of the topological Hubbard model on honeycomb lattice is given
by\cite{Haldane,he1,he2}
\begin{align}
H  &  =-t\sum \limits_{\left \langle {i,j}\right \rangle ,\sigma}\left(  \hat
{c}_{i\sigma}^{\dagger}\hat{c}_{j\sigma}+h.c.\right)  -t^{\prime}%
\sum \limits_{\left \langle \left \langle {i,j}\right \rangle \right \rangle
,\sigma}e^{i\phi_{ij}}\hat{c}_{i\sigma}^{\dagger}\hat{c}_{j\sigma}\label{1}\\
&  -\mu \sum \limits_{i,\sigma}\hat{c}_{i\sigma}^{\dagger}\hat{c}_{i\sigma
}+U\sum \limits_{i}\hat{n}_{i\uparrow}\hat{n}_{i\downarrow}\nonumber \\
&  +\varepsilon \sum \limits_{i\in{A,}\sigma}\hat{c}_{i\sigma}^{\dagger}\hat
{c}_{i\sigma}-\varepsilon \sum \limits_{i\in{B,}\sigma}\hat{c}_{i\sigma
}^{\dagger}\hat{c}_{i\sigma}.\nonumber
\end{align}
$t$ and $t^{\prime}$ are the nearest neighbor and the next nearest neighbor
hoppings, respectively. We introduce a complex phase $\phi_{ij}$ $\left(
\left \vert \phi_{ij}\right \vert =\frac{\pi}{2}\right)  $ to the next nearest
neighbor hopping, of which the positive phase is set to be clockwise. $U$ is
the on-site Coulomb repulsion. $\mu$ is the chemical potential and $\mu=U/2$
at half-filling. $\varepsilon$ denotes an on-site staggered energy and is set
to be $0.15t$.\

In the non-interacting limit $\left(  U=0\right)  ,$ the ground state is a
$Q=2$ topological insulator with quantum anomalous Hall effect (QAH) for
$t^{\prime}>0.0288t$ and a normal band insulator (BI) for $t^{\prime}%
<0.0288t$. At $t^{\prime}=0.0288t,$\ the electron energy gap closes at high
symmetry points in momentum space. As a result, third order topological
quantum phase transition occurs between QAH and BI. See the dispersion of
electrons for $t^{\prime}=0.0288t$ in FIG.1.

When we consider the on-site Coulomb interaction, the ground state can be an
AF SDW order. We have calculated the mean field value of staggered
magnetization $M$ that represents AF SDW order of the topological Hubbard
model from the definition $\langle \hat{c}_{i,\sigma}^{\dagger}\hat
{c}_{i,\sigma}\rangle=\frac{1}{2}(1+(-1)^{i}\sigma M)$ in Ref.\cite{he2}.
Based on the mean field results, the phase diagram has been obtained in FIG.7
in Ref.\cite{he2}. From the phase diagram we get five different quantum
phases: two are non-magnetic states with $M=0$, BI and QAH, three are magnetic
states with $M\neq0$, A-type topological AF SDW state (A-TSDW), B-type
topological AF SDW state (B-TSDW), and trivial AF SDW state.

\begin{figure}[ptb]
\includegraphics[width=0.5\textwidth]{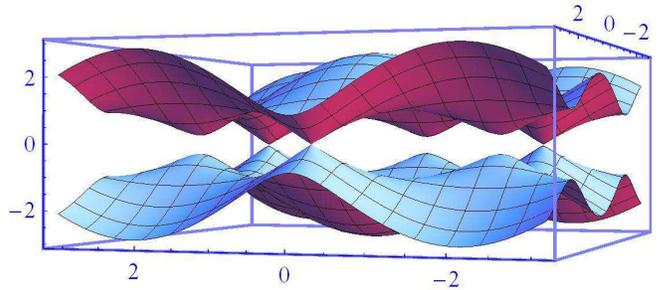}\caption{(Color online) The
dispersion of electrons for $t^{\prime}=0.0288t$ when $U=0.$ We can see
clearly that in the high symmetry point the energy gap is zero and like a
Dirac cone.}%
\end{figure}

Let's explain the quantum phase transitions for different regions of
$t^{\prime}$. For $t^{\prime}>0.0288t,$ the quantum phase transition between a
QAH and AF SDW order is always second order. Thus when we raise the
interaction strength $U,$ due to the smoothly increasing of the staggered
magnetization, the QAH state will turn into the A-TSDW after crossing a
magnetic phase transition, then turn into the B-TSDW crossing a topological
quantum phase transition, eventually turn into the trivial AF SDW state
crossing another topological quantum phase transition. However, in the region
of $t^{\prime}<0.0288t,$ the quantum phase transition between a BI and AF SDW
order is first order which is denoted by the black line in FIG.7 in
Ref.\cite{he2}. Due to the jumping of the staggered magnetization, the BI
state will turn into B-TSDW directly and eventually turn into the trivial AF
SDW state crossing a topological quantum phase transition. In the limit
$t^{\prime}\rightarrow0$, the BI state will change into the trivial AF SDW
state directly and there is no topological state at all. For the case of
$t^{\prime}=0.0288t,$\ it is a semi-metal for the weak coupling limit $\left(
U/t<2.5\right)  $ without electron gap. When we raise the interaction strength
$U,$\ due to the smoothly increasing of the staggered magnetization, the
semi-metal state will turn into the B-TSDW after crossing a magnetic phase
transition, eventually turn into the trivial AF SDW state crossing a
topological quantum phase transition.

\section{Effective $\mathrm{NL}\sigma \mathrm{M}$ for magnetic states}

For the topological Hubbard model on honeycomb lattice, there are three
different magnetic states, A-TSDW, B-TSDW, and trivial AF SDW. A question here
is whether these three SDWs with $M\neq0$ are real long range AF order. The
non-zero value of $M$ by mean field method only means the existence of
effective spin moments. It does not necessarily imply that the ground state is
a long range AF order because the direction of the spins is chosen to be fixed
along $\mathbf{\hat{z}}$-axis in the mean field theory. Thus we will examine
the stability of magnetic order against quantum spin fluctuations of effective
spin moments based on a formulation by keeping spin rotation symmetry,
$\sigma_{z}\rightarrow \mathbf{\Omega}\cdot \mathbf{\sigma.}$

By replacing the electronic operators $\hat{c}_{i}^{\dagger}$ and $\hat{c}%
_{j}$ by Grassmann variables $c_{i}^{\ast}$ and $c_{j}$, in the magnetic
state, we get the effective Lagrangian with spin rotation symmetry as%
\begin{align}
\mathcal{L}_{\mathrm{eff}}  &  =\sum_{i,\sigma}c_{i\sigma}^{\ast}%
\partial_{\tau}c_{i\sigma}-t\sum \limits_{\left \langle {i,j}\right \rangle
,\sigma}\left(  c_{i\sigma}^{\ast}c_{j\sigma}+h.c.\right) \nonumber \\
&  -t^{\prime}\sum \limits_{\left \langle \left \langle {i,j}\right \rangle
\right \rangle ,\sigma}e^{i\phi_{ij}}c_{i\sigma}^{\ast}c_{j\sigma}-\sum
_{i}\left(  -1\right)  ^{i}\Delta_{M}c_{i\sigma}^{\ast}\mathbf{\Omega}%
_{i}\mathbf{\cdot \sigma}c_{i\sigma}\nonumber \\
&  +\varepsilon \sum \limits_{i\in{A,}\sigma}c_{i\sigma}^{\ast}c_{i\sigma
}-\varepsilon \sum \limits_{i\in{B,}\sigma}c_{i\sigma}^{\ast}c_{i\sigma}.
\end{align}
Where $\Delta_{M}=UM/2$, Within the Haldane's mapping, the spins are
parametrized as $\mathbf{\Omega}_{i}=(-1)^{i}\mathbf{n}_{i}\sqrt
{1-\mathbf{L}_{i}^{2}}+\mathbf{L}_{i}$%
\cite{Haldane1,Auerbach,dup,dupuis,Schulz,weng}. Here $\mathbf{n}_{i}$ is the
N\'{e}er vector and $\left \vert \mathbf{n}_{i}\right \vert =1$, $\mathbf{L}%
_{i}$ is the transverse canting field, which is chosen to $\mathbf{L}_{i}%
\cdot \mathbf{n}_{i}=0.$

Then we integrate fermions and the transverse canting field and obtain the
effective $\mathrm{NL}\sigma \mathrm{M}$ as
\begin{equation}
\mathcal{L}_{\mathbf{n}}=\frac{1}{2g}[\frac{1}{c}\left(  \partial_{\tau
}\mathbf{n}\right)  ^{2}+c\left(  \mathbf{\bigtriangledown n}\right)
^{2}]\text{ }%
\end{equation}
with a constraint $\mathbf{n}^{2}=1.$ The coupling constant $g$ and spin wave
velocity $c$ are defined as
\begin{equation}
g=\frac{c}{\rho_{s}},\text{ }c^{2}=\frac{\rho_{s}}{\chi^{\perp}}.
\end{equation}
Here $\rho_{s}$ is the spin stiffness and $\chi^{\perp}$ is the transverse
spin susceptibility.\ The detailed calculations are given in Appendix. A.

\begin{figure}[ptb]
\includegraphics[width=0.5\textwidth]{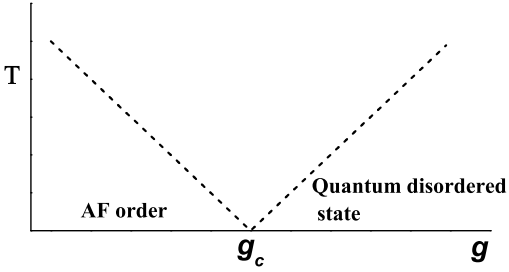}\caption{The illustration of
the relationship between AF order and quantum disordered state}%
\end{figure}

The properties of the effective $\mathrm{NL}\sigma \mathrm{M}$ are determined
by the dimensionless coupling constant $\alpha=g\Lambda.$ The cutoff is
defined as the following equation $\Lambda=\min(1,\Delta E/c).$ Here $\Delta
E$ is the energy gap of electrons. In particular, there exists a critical
point $\alpha_{c}=4\pi$ (or $g_{c}=\frac{4\pi}{\Lambda}$). See illustration of
FIG.2. The quantum critical point (QCP) separates the long range spin order
from the short range spin order (the quantum disordered state). The dotted
line shows the renormalized spin stiffness of the long range spin order and
the energy scale of spin gap of the quantum disordered state, respectively
(see below discussion).

For the case of $\alpha<4\pi,$ we get solutions of the spin condensed $n_{0}$
and spin gap $m_{s}$ at zero temperature:%
\begin{equation}
n_{0}=(1-\frac{g}{g_{c}})^{1/2}\text{, }m_{s}=0.
\end{equation}
At finite temperature, the solutions become $n_{0}=0$ and $m_{s}%
=2k_{\mathrm{B}}T\sinh^{-1}[e^{-\frac{2\pi c}{gk_{\mathrm{B}}T}}\sinh
(\frac{c\Lambda}{2k_{\mathrm{B}}T})]$. Because the energy scale of the spin
gap $m_{s}$ is always much smaller than the temperature, \textit{i.e.},
$m_{s}\ll k_{\mathrm{B}}T$ (or $\omega_{n}$), quantum fluctuations become
negligible in a sufficiently long wavelength and low energy regime $\left(
m_{s}<\left \vert c\mathbf{q}\right \vert <k_{\mathrm{B}}T\right)  .$ Thus in
this region one may only consider the purely static (semiclassical)
fluctuations. The effective Lagrangian of the NL$\sigma$M then becomes%
\begin{equation}
\mathcal{L}=\frac{\rho_{\mathrm{s,eff}}}{2}\left(  \mathbf{\bigtriangledown
n}\right)  ^{2}%
\end{equation}
where $\rho_{\mathrm{s,eff}}=c\left(  \frac{1}{g}-\frac{1}{g_{c}}\right)  $ is
the renomalized spin stiffness. At zero temperature, the mass gap vanishes
which means that long range AF order appears. To describe the long range AF
order, we introduce a spin order parameter
\begin{equation}
\mathcal{M}_{0}=\frac{M}{2}n_{0}=\frac{M}{2}(1-\frac{g}{g_{c}})^{1/2}\text{,
}m_{s}=0.
\end{equation}
The ground state of long range AF ordered phase has a finite spin order
parameter. And in this region there are two transverse Goldstone modes,
between them the interaction is irrelevant.

For the case of $\alpha>4\pi,$ the interaction between Goldstone modes becomes
relevant and at low energy the renormalized coupling constant diverges.
Consequently, the spin gap opens and the long range spin order disappears
which mean that the ground state may be a quantum disordered state, and we get
the effective model of massive spin-1 excitations
\begin{equation}
\mathcal{L}_{\mathbf{s}}=\frac{1}{2g}\left[  (\partial_{\mu}\mathbf{n}%
)^{2}+m_{s}^{2}\mathbf{n}^{2}\right]
\end{equation}
with the solutions of $n_{0}$ and $m_{s}$ as%
\begin{equation}
n_{0}=0\text{, }m_{s}=4\pi c(\frac{1}{g_{c}}-\frac{1}{g}).
\end{equation}
Using the CP(1) representation, we have
\begin{equation}
\mathcal{L}_{\mathbf{s}}=\frac{2}{g}\left[  |(\partial_{\mu}-ia_{\mu
})\mathbf{z}|^{2}+m_{z}^{2}\mathbf{z}^{2}\right]
\end{equation}
where $\mathbf{z}$ is a bosonic spinon, $\mathbf{z}=\left(  z_{1},\text{
}z_{2}\right)  ^{T},$ $\mathbf{n}_{i}=\mathbf{\bar{z}}_{i}\mathbf{\sigma
z}_{i}\mathbf{,}$ $\mathbf{\bar{z}z=1,}$ $a_{\mu}\equiv-\frac{i}%
{2}(\mathbf{\bar{z}}\partial_{\mu}\mathbf{z}-\partial_{\mu}\mathbf{\bar{z}z}%
)$. Here $a_{\mu}$\ is introduced as an assistant gauge field. Specifically
the local gauge transformation is $z\rightarrow e^{i\varphi(r,\tau)}z$.
$m_{z}$ denotes the mass gap for spinons as $m_{z}=m_{s}/2$.

In addition, after integrating over fermions by using gradient expansion
approach we also obtain the Chern-Simons-Hopf (CSH) term as\cite{Kmat,he2}
\begin{equation}
\mathcal{L}_{CSH}=-i\sum_{I,J}\frac{\mathcal{K}_{IJ}}{4\pi}\varepsilon^{\mu
\nu \lambda}a_{\mu}^{I}\partial_{\nu}a_{\lambda}^{J}%
\end{equation}
where $\mathcal{K}$ is 2-by-2 matrix, $a_{\mu}^{I=1}=A_{\mu}$ and $a_{\mu
}^{I=2}=a_{\mu}.$ $A_{\mu}$ is the electric-magnetic field. The "charge" of
$A_{\mu}$ and $a_{\mu}$ are defined by $q$ and $q_{s}$, respectively. Thus for
different SDW orders with the same order parameter $M$, we have different
$\mathcal{K}$-matrices : for A-TSDW order, $\mathcal{K}=\left(
\begin{array}
[c]{ll}%
2 & 0\\
0 & 2
\end{array}
\right)  ;$ for B -TSDW order, $\mathcal{K}=\left(
\begin{array}
[c]{ll}%
1 & 1\\
1 & 1
\end{array}
\right)  ;$ for trivial SDW order, $\mathcal{K}=0.$ See detailed calculations
in Appendix. B.

\begin{figure}[ptb]
\includegraphics[width=0.5\textwidth]{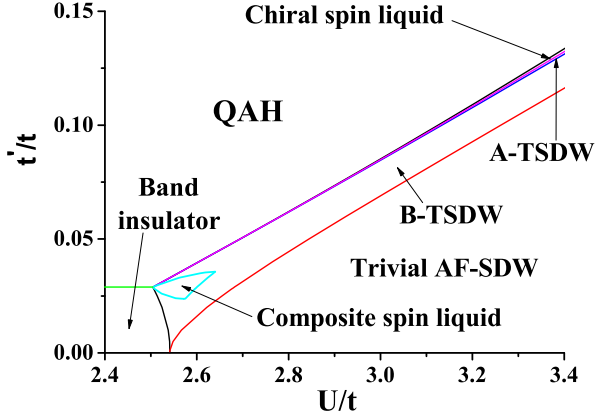}\caption{(Color online) The
phase diagram: there are seven phases, QAH, band insulator, A-TSDW, B-TSDW,
chiral-spin-liquid, composite spin liquid and trivial AF-SDW. The regions of
chiral spin liquid and composite spin liquid are the quantum disordered
regions of $\alpha>4\pi$.}%
\end{figure}

\begin{figure}[ptb]
\includegraphics[width=0.5\textwidth]{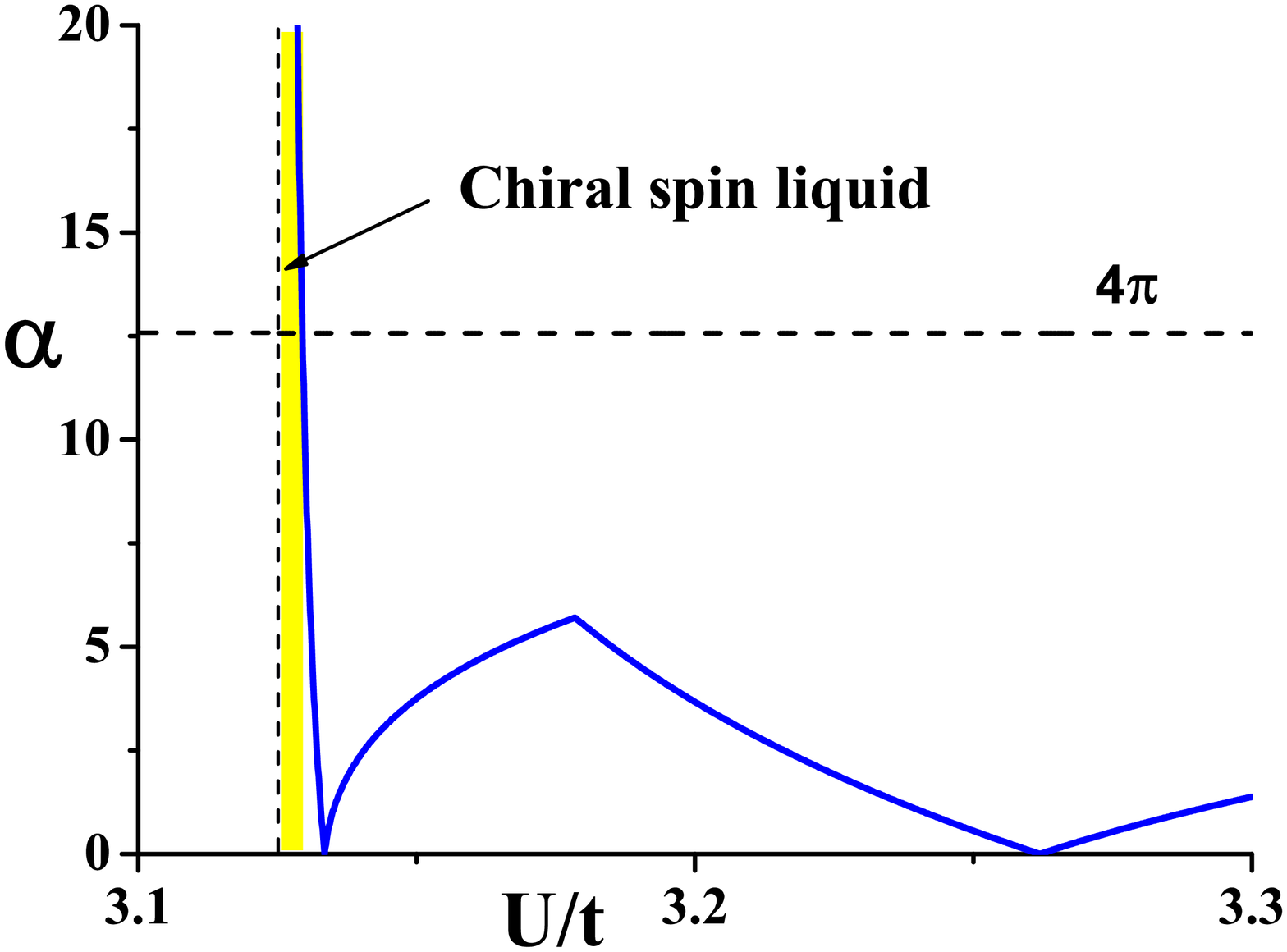}\caption{(Color online) The
dimensionless coupling constant $\alpha=g\Lambda$ for the case of the
parameter as $t^{\prime}=0.1t.$ For the region with $\alpha>4\pi$, the ground
state is chiral spin liquid (yellow region).}%
\end{figure}

\begin{figure}[ptb]
\includegraphics[width=0.5\textwidth]{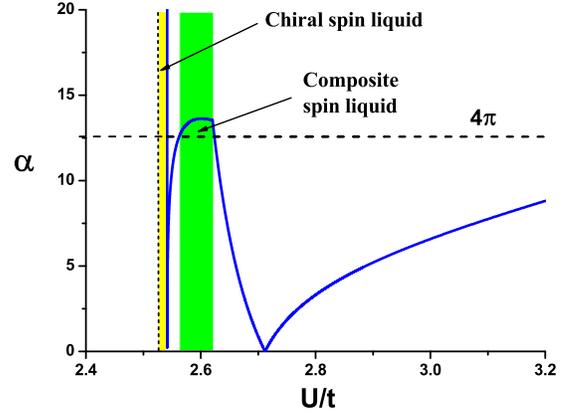}\caption{(Color online) The
dimensionless coupling constant $\alpha=g\Lambda$ for the case of the
parameter as $t^{\prime}=0.033t.$ For the regions with $\alpha>4\pi$, the
ground states are spin liquid states - chiral spin liquid (yellow region) or
composite spin liquid (green region).}%
\end{figure}

\begin{figure}[ptb]
\includegraphics[width=0.5\textwidth]{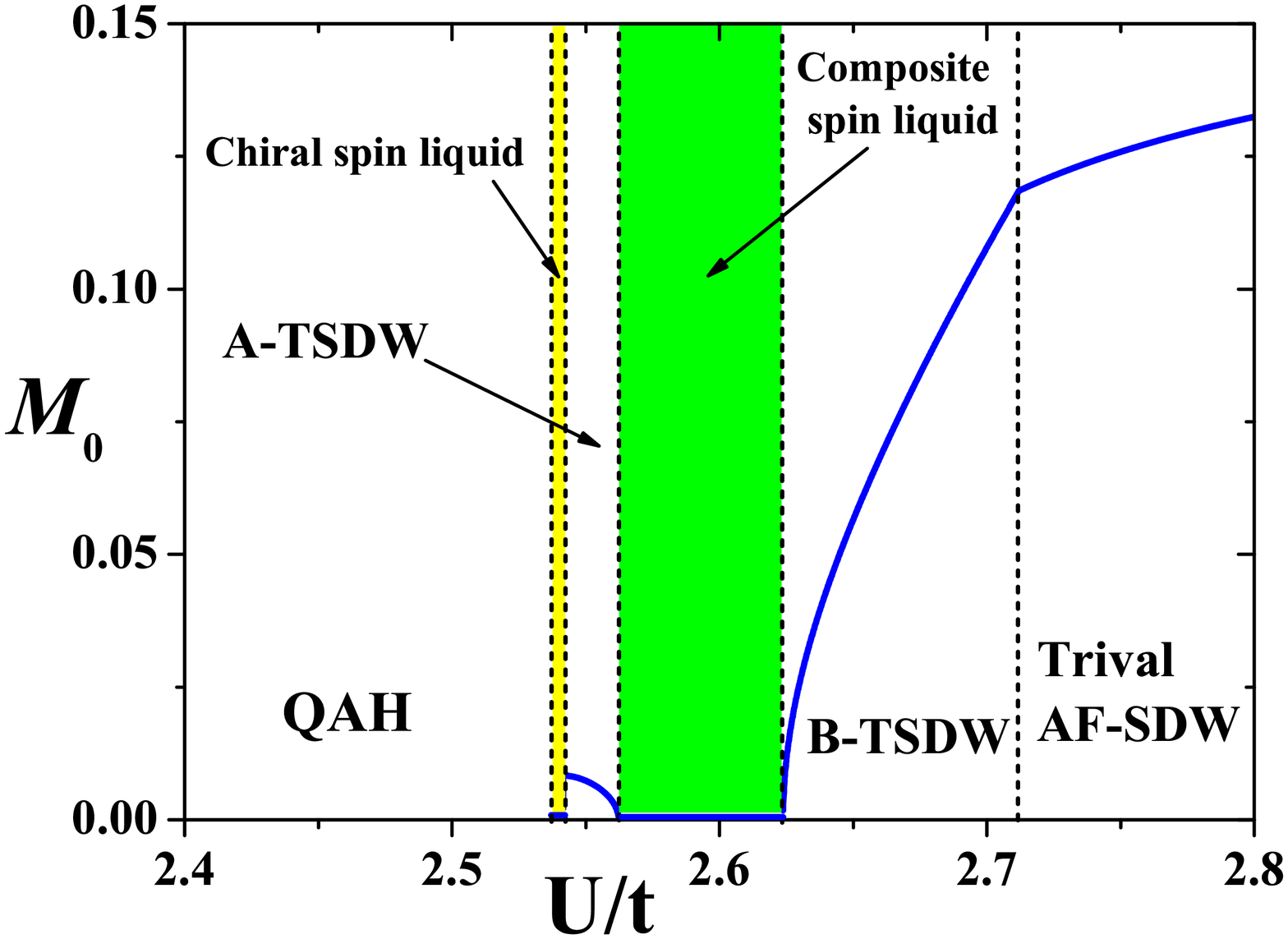}\caption{(Color online) The
spin order parameter $\mathcal{M}_{0}$ for the case of the parameter as
$t^{\prime}=0.033t$. Yellow region denotes chiral spin liquid and green region
denotes composite spin liquid, of which $\mathcal{M}_{0}=0$.}%
\end{figure}

\begin{figure}[ptb]
\includegraphics[width=0.5\textwidth]{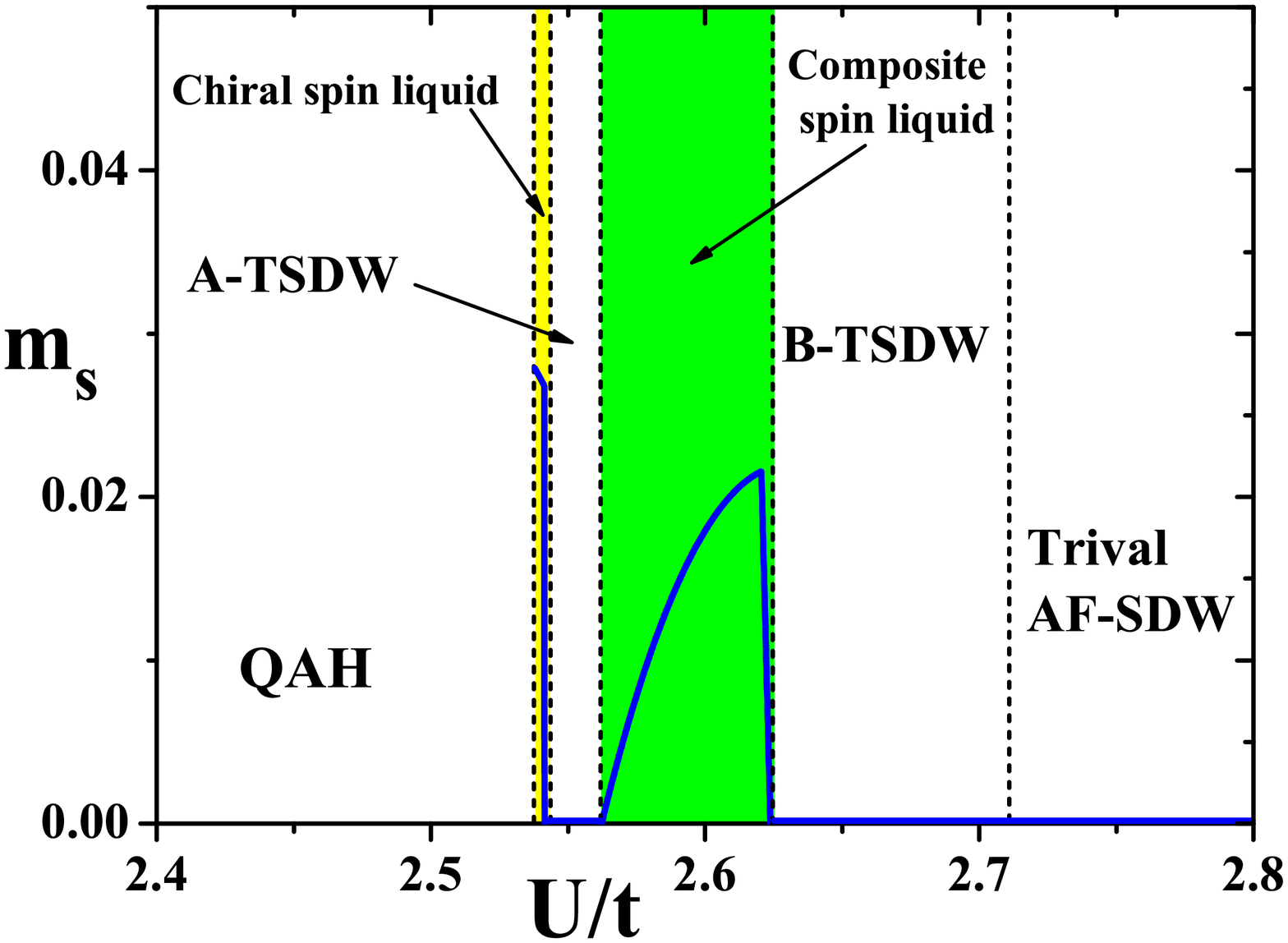}\caption{(Color online) The
spin gap $m_{s}$ for the case of the parameter as $t^{\prime}=0.033t.$ Yellow
region denotes chiral spin liquid and green region denotes composite spin
liquid, of which $m_{s}\neq0$.}%
\end{figure}

\begin{figure}[ptb]
\includegraphics[width=0.5\textwidth]{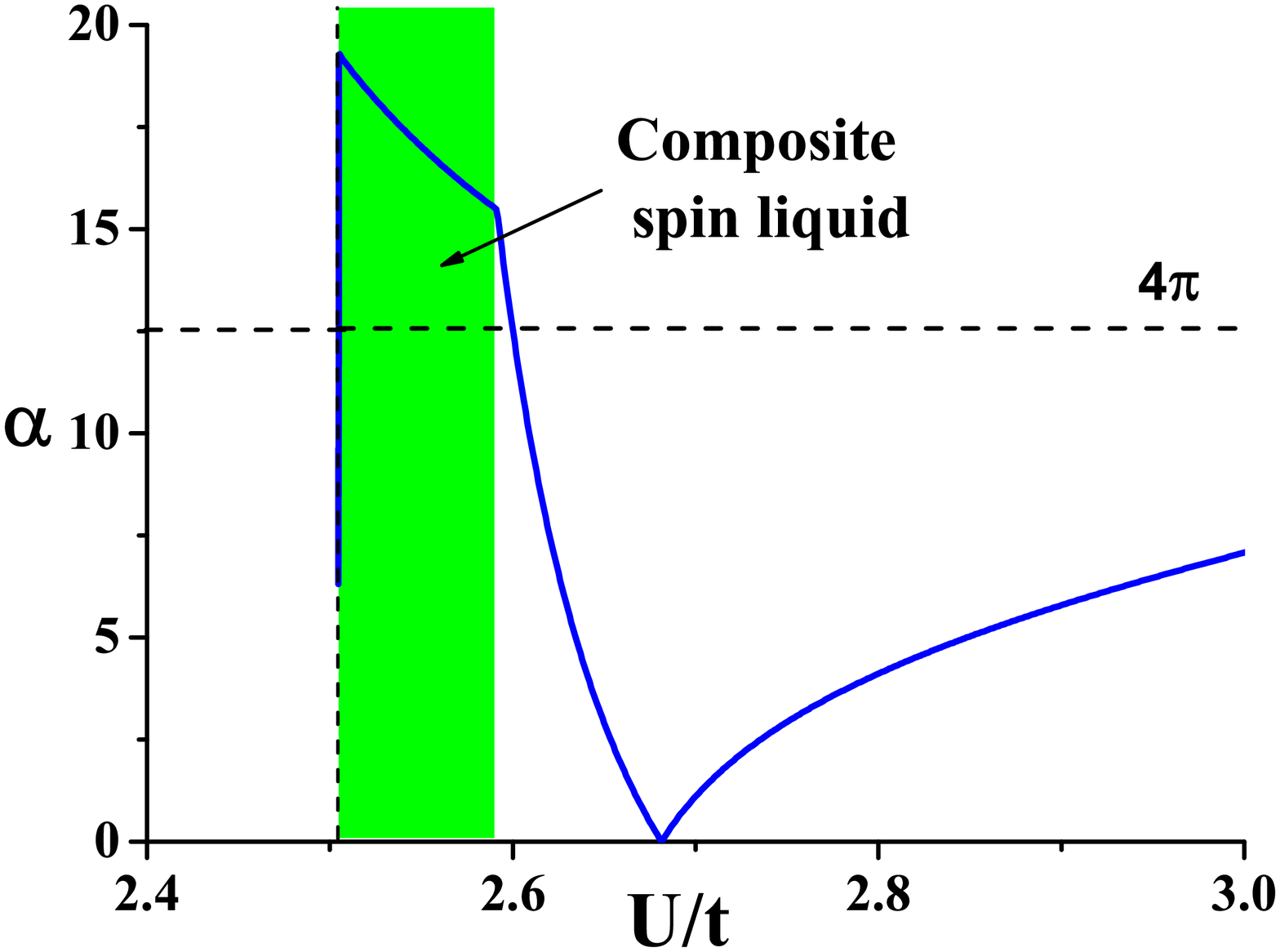}\caption{(Color online) The
dimensionless coupling constant $\alpha=g\Lambda$ for the case of the
parameter as $t^{\prime}=0.0288t.$ For the region with $\alpha>4\pi$, the
ground state is composite spin liquid (green region).}%
\end{figure}

\begin{figure}[ptb]
\includegraphics[width=0.5\textwidth]{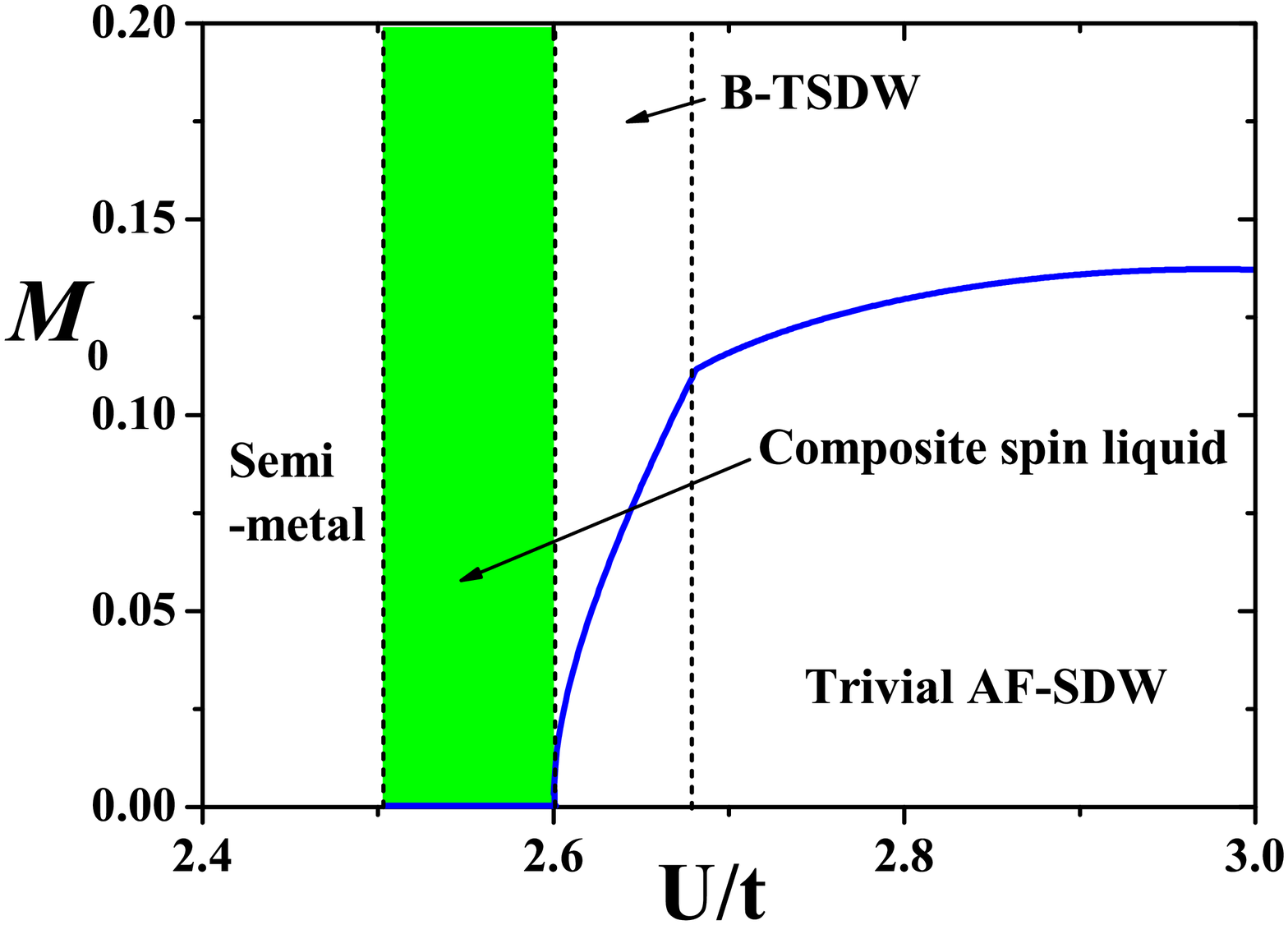}\caption{(Color online) The
spin order parameter $\mathcal{M}_{0}$ for the case of the parameter as
$t^{\prime}=0.0288t$. The green region denotes composite spin liquid, of which
$\mathcal{M}_{0}=0$.}%
\end{figure}

\begin{figure}[ptb]
\includegraphics[width=0.5\textwidth]{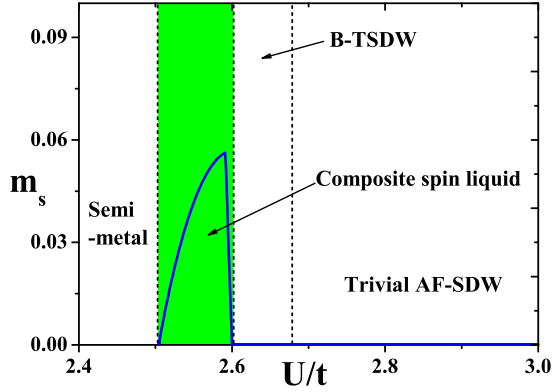}\caption{(Color online) The
spin gap $m_{s}$ for the case of the parameter as $t^{\prime}=0.0288t.$ The
green region denotes composite spin liquid, of which $m_{s}\neq0$.}%
\end{figure}

\begin{figure}[ptb]
\includegraphics[width=0.5\textwidth]{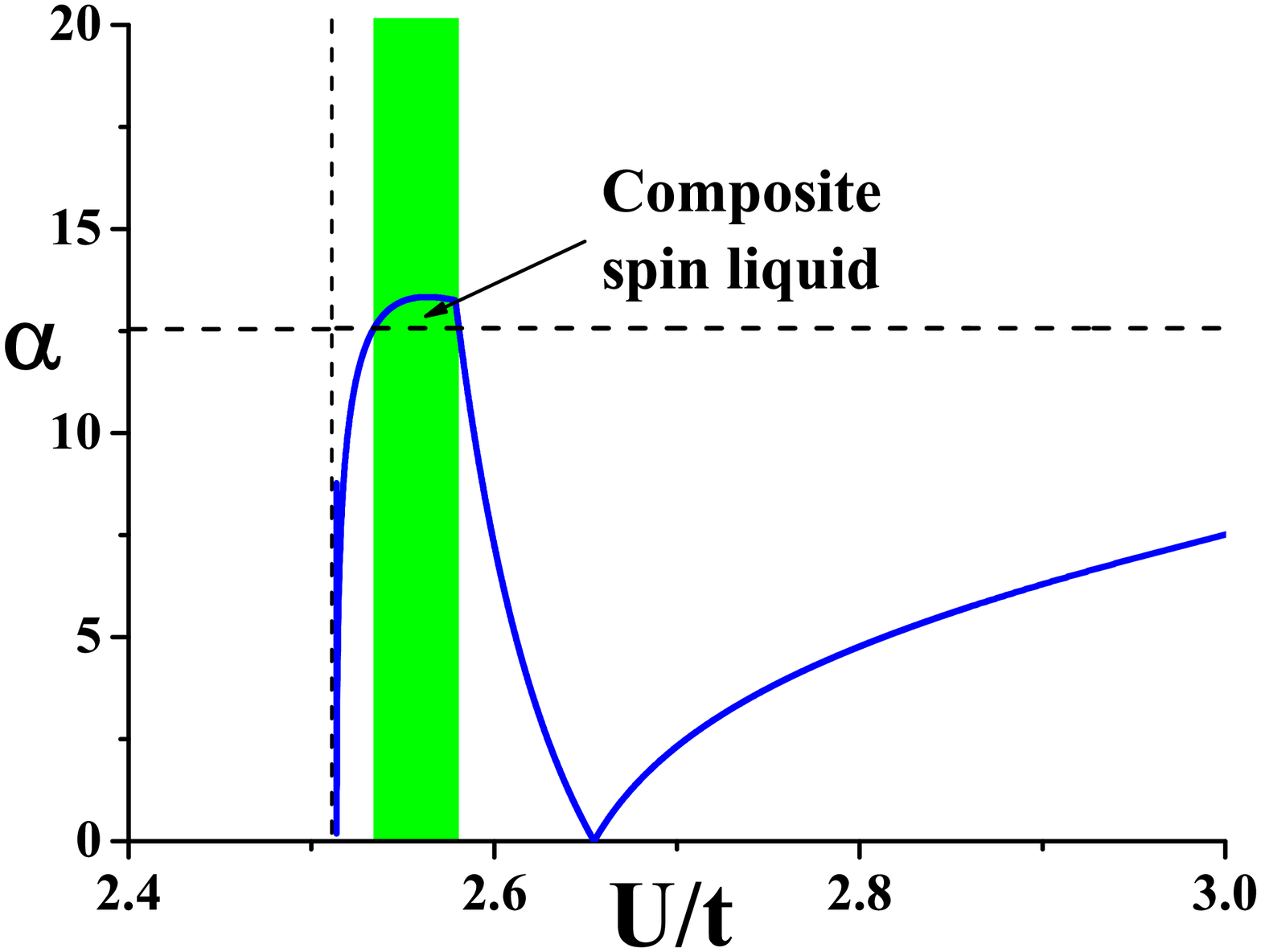}\caption{(Color online) The
dimensionless coupling constant $\alpha=g\Lambda$ for the case of the
parameter as $t^{\prime}=0.025t.$ For the region with $\alpha>4\pi$, the
ground state is composite spin liquid (green region).}%
\end{figure}

\begin{figure}[ptb]
\includegraphics[width=0.5\textwidth]{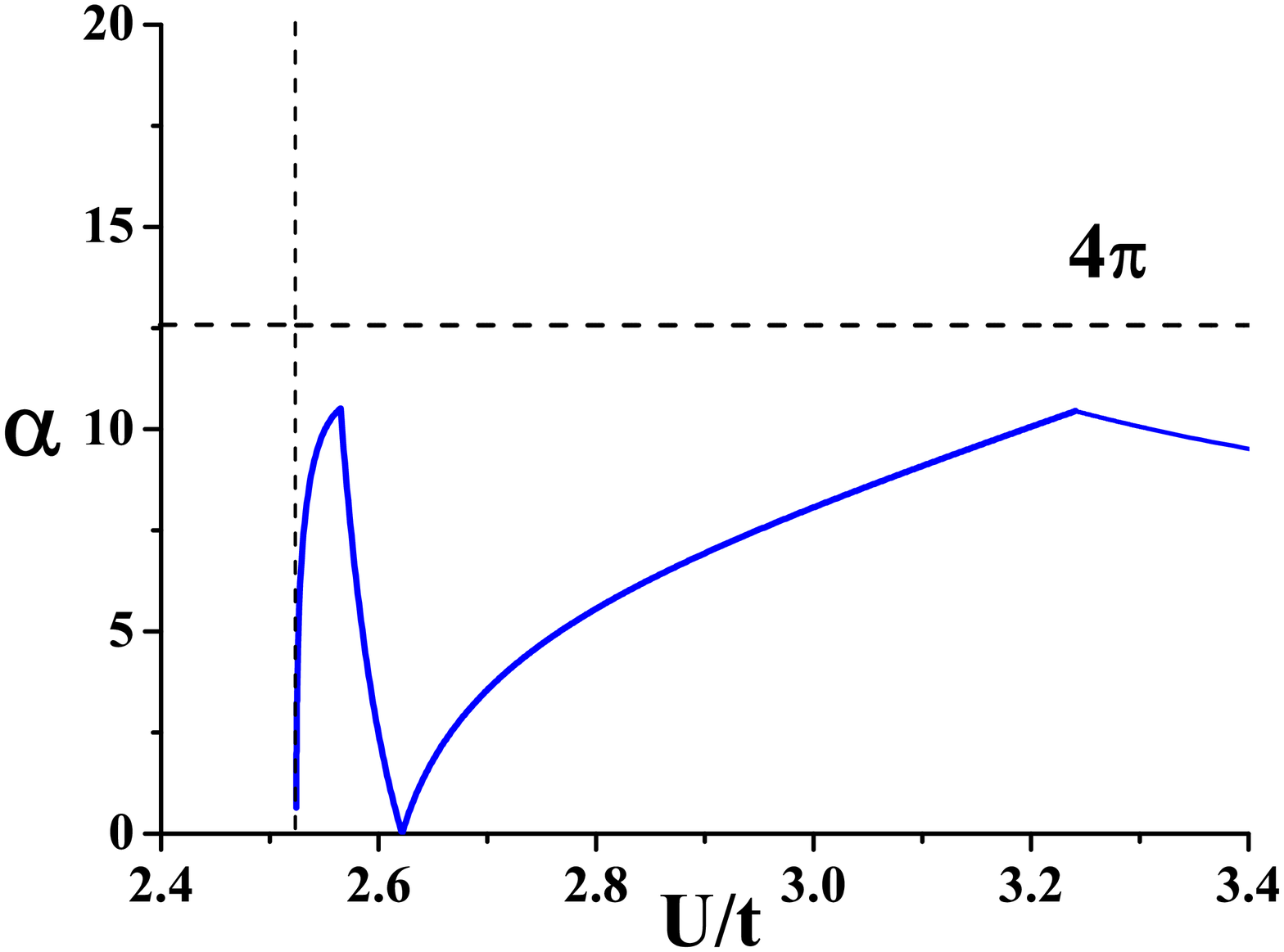}\caption{(Color online) The
dimensionless coupling constant $\alpha=g\Lambda$ for the case of the
parameter as $t^{\prime}=0.02t.$ We can see that the dimensionless coupling
constant $\alpha$ is always smaller than $\alpha_{c}=4\pi.$ That mean there
doesn't exist quantum disordered region at all.}%
\end{figure}

For different regions of $t^{\prime},$ we calculated the dimensionless
coupling constant $g$ ($\alpha$) and derived the quantum phase transitions
between long range AF SDW order and short range one. Thus we can plot a new
phase diagram in FIG.3 that shows the quantum disordered regions of
$\alpha>4\pi$ (The regions of chiral spin liquid and composite spin liquid).

For a given $t^{\prime}$ bigger than $0.0288t,$ there are two situations.
FIG.4 shows the dimensionless coupling constant for one situation with the
parameter $t^{\prime}=0.1t$. In FIG.4 there exists a quantum disordered region
with $\alpha>\alpha_{c}=4\pi$ in A-TSDW that corresponds to the chiral spin
liquid (yellow region). The other case is shown in FIG.5, of which the
dimensionless coupling constant for the parameter $t^{\prime}=0.033t$. There
are two quantum disordered regions with $\alpha>\alpha_{c}=4\pi$: one
corresponds to the chiral spin liquid (yellow region) in A-TSDW, the other is
composite spin liquid (green region) in B-TSDW (see discussion in following
sections). For this case, we get the energy gap of spin order parameter
$\mathcal{M}_{0}$ and spin excitations $m_{s}$ in FIG.6 and FIG.7. One can see
that in chiral spin liquid and composite spin liquid, $\mathcal{M}_{0}=0$,
$m_{s}\neq0$. For the case of $t^{\prime}=0.0288t,$\ we show the result of the
dimensionless coupling constant in FIG.8, from which one can see that there
exists a quantum disordered region with $\alpha>\alpha_{c}=4\pi$ in B-TSDW
that corresponds to the composite spin liquid (green region). For this case,
we also get the energy gap of spin order parameter $\mathcal{M}_{0}$ and spin
excitations $m_{s}$ in FIG.9 and FIG.10. One can see that in composite spin
liquid, $\mathcal{M}_{0}=0$, $m_{s}\neq0$. For a given $t^{\prime}$ smaller
than $0.0288t,$ there are also two situations. For $0.02377t<t^{\prime
}<0.0288t$ $\left(  t^{\prime}=0.025t\right)  ,$ from the result shown in
FIG.11, we found a quantum disordered region with $\alpha>\alpha_{c}=4\pi$ in
B-TSDW that corresponds to the composite spin liquid (green region). For
$0<t^{\prime}<0.02377t$ $\left(  t^{\prime}=0.02t\right)  ,$ we found that the
dimensionless coupling constant $\alpha$ is always smaller than $\alpha
_{c}=4\pi.$ That mean there doesn't exist quantum disordered region at all. We
also plot FIG.12 to show this situation.

In the following parts we will use the effective model with CSH terms to learn
the properties of different SDW orders\cite{he2},
\[
\mathcal{L}_{\mathrm{eff}}=\mathcal{L}_{\mathbf{s}}+\mathcal{L}_{CSH}%
\]
where
\[
\mathcal{L}_{\mathbf{s}}=\frac{1}{2g}\left[  (\partial_{\mu}\mathbf{n}%
)^{2}+m_{s}^{2}\mathbf{n}^{2}\right]
\]
and
\[
\mathcal{L}_{CSH}=\sum_{I,J}\frac{K_{IJ}}{4\pi}\varepsilon^{\mu \nu \lambda
}a_{\mu}^{I}\partial_{\nu}a_{\lambda}^{J}.
\]

Thus an important issue is that \emph{what's the nature of these quantum
disordered states with different CSH terms}. Our answer is : for the case of
A-TSDW with $\mathcal{K}=\left(
\begin{array}
[c]{ll}%
2 & 0\\
0 & 2
\end{array}
\right)  ,$ the quantum disordered state is a chiral spin liquid with
topological degeneracy and anyonic excitations (See illustration of FIG.13);
for the case of B-TSDW with $\mathcal{K}=\left(
\begin{array}
[c]{ll}%
1 & 1\\
1 & 1
\end{array}
\right)  ,$ the quantum disordered state is composite spin liquid with chiral
edge states, of which the elementary excitation is spin one-half charge $\pm
e$ objects trapping a topological spin texture (See illustration of FIG.15).

\section{Chiral spin liquid - quantum disordered state of A-TSDW}

\begin{figure}[ptb]
\includegraphics[width=0.5\textwidth]{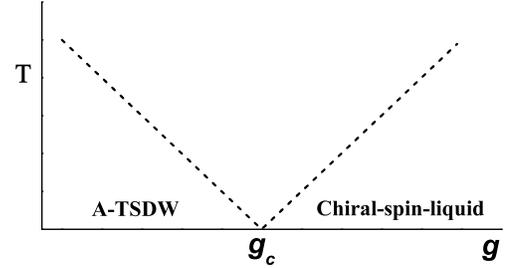}\caption{The illustration of
the relationship between A-TSDW and chiral spin liquid}%
\end{figure}

Firstly, we study the quantum disordered state of A-TSDW that is described by
\begin{align*}
\mathcal{L}_{\mathrm{eff}}  &  =\frac{1}{2g}\left[  (\partial_{\mu}%
\mathbf{n})^{2}+m_{s}^{2}\mathbf{n}^{2}\right] \\
&  +\frac{1}{2\pi}\epsilon^{\mu \nu \lambda}A_{\mu}\partial_{\nu}A_{\lambda
}+\frac{1}{2\pi}\epsilon^{\mu \nu \lambda}a_{\mu}\partial_{\nu}a_{\lambda},
\end{align*}
or%
\begin{align}
\mathcal{L}_{\mathrm{eff}}  &  =\frac{2}{g}\left[  |(\partial_{\mu}-ia_{\mu
})\mathbf{z}|^{2}+m_{z}^{2}\mathbf{z}^{2}\right] \label{a}\\
&  +\frac{1}{2\pi}\epsilon^{\mu \nu \lambda}A_{\mu}\partial_{\nu}A_{\lambda
}+\frac{1}{2\pi}\epsilon^{\mu \nu \lambda}a_{\mu}\partial_{\nu}a_{\lambda
}.\nonumber
\end{align}

At low energy limit, the kinetic term of gauge field $a_{\mu}$ is induced%
\begin{equation}
\mathcal{L}\left(  a_{\mu}\right)  =\frac{1}{4e_{a}^{2}}\left(  \partial_{\mu
}a_{\nu}\right)  ^{2}.
\end{equation}
The induced coupling constant of three dimensional gauge field is $e_{a}%
^{2}=3\pi m_{z}^{2}$. After considering the CSH term, we have the effective
Lagrangian as
\begin{align}
\mathcal{L}_{\mathrm{eff}}  &  =\frac{1}{4e_{a}^{2}}(\partial_{\mu}a_{\nu
})^{2}+\frac{1}{2\pi}\epsilon^{\mu \nu \lambda}a_{\mu}\partial_{\nu}a_{\lambda
}\nonumber \\
&  +\frac{1}{2\pi}\epsilon^{\mu \nu \lambda}A_{\mu}\partial_{\nu}A_{\lambda}.
\end{align}

For the compact \textrm{U(1)} gauge theory in 2+1 dimensions, there exist the
instantons (space-time `magnetic' monopoles) that generate $2\pi$ gauge flux
of $a_{\mu}$ indicates that $a_{\mu}$ gauge field is `compact'\cite{confine}.
Without the CSH term, the monopoles form Coulomb gas in 2+1 dimensions. Due to
the Debye screening in the monopole plasma, the gauge field $a_{\mu}$ obtains
a mass gap and bosonic spinons $\mathbf{z}$ that couple the gauge field
$a_{\mu}$ are confined. And it is pointed out in Ref.\cite{wen3} that from the
Berry phase of path integral of spin coherent state on honeycomb lattice, the
ground state with spinon-confinement is really a VBS\ state with spontaneous
translation symmetry breaking.

However, due to the Chern-Simon term, $\frac{1}{2\pi}\epsilon^{\mu \nu \lambda
}a_{\mu}\partial_{\nu}a_{\lambda},$ the instantons are confined by linear
potential and irrelevant to low energy physics. Thus the ground state cannot
be VBS state and spinons are deconfined. In particular, the Chern-Simons term
for $a_{\mu}$ has a nontrivial statistics effect. Because the low energy
physics is dominated only by spinon $\mathbf{z}$, due to $\frac{1}{2\pi
}\epsilon^{\mu \nu \lambda}a_{\mu}\partial_{\nu}a_{\lambda}$, the statistics
angel of $\mathbf{z}$ is $\pi/2$. As a result, spinons $\mathbf{z}$ becomes a
semionic particle with spin $J=\frac{1}{4}$! Therefore the quantum disordered
state of A-TSDW that is described by the effective Lagrangian in Eq.[\ref{a}]
is really a topological ordered state - \emph{chiral spin liquid}. From the
CSH term, one may derive topological degeneracy - two degenerate ground states
of chiral spin liquid on a torus\cite{chiral}. The result is consistent to
that in Ref.\cite{he1}.

In addition, one can also derive the edge states from the effective CSH
theory. There are two right-moving "spin" edge excitations described by the
following 1D fermion theory\cite{edge}
\[
\mathcal{L}_{\text{\textrm{edge}}}=\sum_{\alpha}\psi_{\alpha s}^{\dag
}(\partial_{t}-v_{R}\partial_{x})\psi_{\alpha s},
\]
where $\alpha=1,2.$ $\psi_{\alpha s}$ carries a unit of $a_{\mu}$ charge. One
can see "spin" chiral edge states in FIG.14 (the lines with arrows).
Correspondingly, one can get the quantized spin Hall conductivity
\begin{equation}
\sigma_{s}=\lim_{\omega \rightarrow0}{\frac{1}{\omega}}\epsilon_{ij}%
\left \langle J_{si}(\omega,0)J_{sj}(-\omega,0)\right \rangle =\frac{2e^{2}}{h}.
\end{equation}
Here $J_{si}$ denotes spin current, $J_{si}=-i\langle \sum_{a}\bar{\psi}%
_{a}\gamma_{i}\mathbf{n\cdot \sigma}\psi_{a}\rangle$.

On the other hand, we discuss the properties of $A_{\mu}$. The gauge field
$A_{\mu}$ is classical field and has no dynamic terms. Thus the Chern-Simons
term for $A_{\mu}$ only indicates quantized anomalous charge Hall effect. From
it, we find two right-moving branches of "charge" edge excitations, which are
described by the following one dimension fermion theory\cite{edge0}
\begin{equation}
\mathcal{L}_{\text{\textrm{edge}}}=\sum_{\alpha}\psi_{c,\alpha}^{\dag
}(\partial_{t}-v_{c}\partial_{x})\psi_{c,\alpha},
\end{equation}
where $\alpha,\beta=1,2.$ $\psi_{c,\alpha}$ carries a unit of $A_{\mu}$
charge. One can see "charge" chiral edge state (the lines with dots) in
FIG.14. Consequently, we get the quantized charge Hall conductivity
$\sigma_{H}=\frac{2e^{2}}{h}.$

\begin{figure}[ptb]
\includegraphics[width=0.5\textwidth]{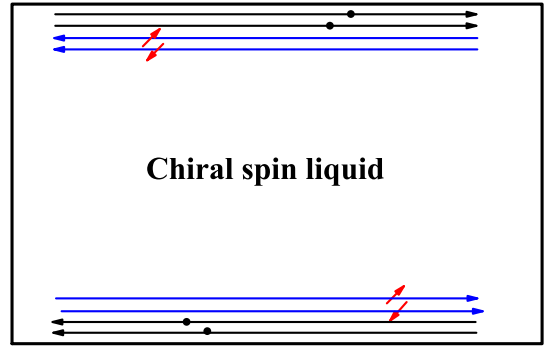}\caption{(Color online) The
illustration of the edge state of chiral spin liquid. There exist "spin"
chiral edge state (the lines with arrows) and "charge" chiral edge state (the
lines with dots).}%
\end{figure}

Finally, we identify the quantum disordered state of A-TSDW characterized by
$g>g_{c}$ to be a chiral spin liquid with quantum anomalous Hall effect (See
illustration of FIG.13). For this system, there exists spin-charge separation.
In FIG.13, the QCP at $g=g_{c}$ denotes the quantum phase transition dividing
long range A-TSDW and short range A-TSDW (chiral SL). In addition, we should
emphasis the existence of the chiral spin liquid due to strongly fluctuated
spin moments characterized by the diverge behavior of the spin coupling
constant near the quantum phase transition (yellow region) in FIG.4 and FIG.5
as $g\rightarrow g_{c}$. Thus the existence of the chiral spin liquid is
independent on the cutoff $\Lambda$.

\section{Composite spin liquid - quantum disordered state of B-TSDW}

\begin{figure}[ptb]
\includegraphics[width=0.5\textwidth]{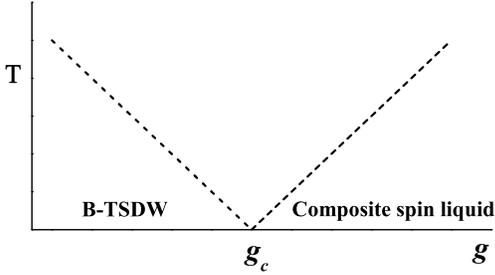}\caption{The illustration of
the relationship between B-TSDW and composite spin liquid. }%
\end{figure}

Next we study the quantum disordered state of B-TSDW that is described by the
low energy effective Lagrangian
\begin{align*}
\mathcal{L}_{\mathrm{eff}}  &  =\frac{1}{2g}\left[  (\partial_{\mu}%
\mathbf{n})^{2}+m_{s}^{2}\mathbf{n}^{2}\right]  +\frac{1}{4\pi}\epsilon
^{\mu \nu \lambda}A_{\mu}\partial_{\nu}A_{\lambda}\\
&  +\frac{1}{2\pi}\epsilon^{\mu \nu \lambda}A_{\mu}\partial_{\nu}a_{\lambda
}+\frac{1}{4\pi}\epsilon^{\mu \nu \lambda}a_{\mu}\partial_{\nu}a_{\lambda}%
\end{align*}
or
\begin{align*}
\mathcal{L}_{\mathrm{eff}}  &  =\frac{2}{g}\left[  |(\partial_{\mu}-ia_{\mu
})\mathbf{z}|^{2}+m_{z}^{2}\mathbf{z}^{2}\right] \\
&  +\frac{1}{2\pi}\epsilon^{\mu \nu \lambda}A_{\mu}\partial_{\nu}a_{\lambda
}+\frac{1}{4\pi}\epsilon^{\mu \nu \lambda}A_{\mu}\partial_{\nu}A_{\lambda}%
+\frac{1}{4\pi}\epsilon^{\mu \nu \lambda}a_{\mu}\partial_{\nu}a_{\lambda}.
\end{align*}
From FIG.15, one can find that there indeed exist a region of short range
B-TSDW order that is characterized by $g>g_{c}$.

\begin{figure}[ptb]
\includegraphics[width=0.5\textwidth]{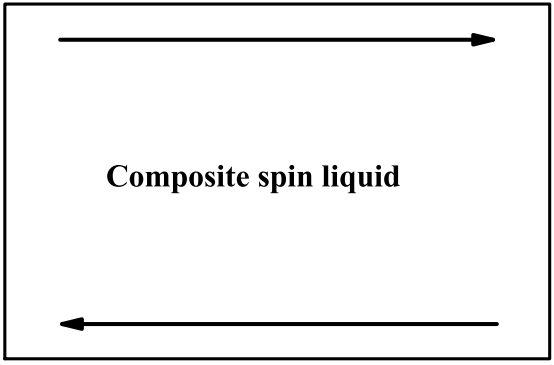}\caption{The illustration of
the edge state of composite spin liquid. There exists a single chiral edge
mode.}%
\end{figure}

Firstly, we study the statistics of spinon $\mathbf{z}$. To learn the
statistics of spinon $\mathbf{z,}$ we can set $A_{\mu}$ to be zero due to
$A_{\mu}$ is a classical field. Thus the CS term is reduced into $\frac
{1}{4\pi}\epsilon^{\mu \nu \lambda}a_{\mu}\partial_{\nu}a_{\lambda}$. From it we
can see that the spinons are fermionic particle by binding a $2\pi$ flux of
$a_{\mu}$ that is just a skyrmion\ (or an anti-skyrmion). On the other hand,
due to the mutual Chern-Simons term $\frac{1}{2\pi}\epsilon^{\mu \nu \lambda
}A_{\mu}\partial_{\nu}a_{\lambda}$, a $2\pi$ flux of $a_{\mu}$ will carry a
electric charge. Thus $\mathbf{z}$ particle is really an "electron" or a
"hole" binding a skyrmion\ (or anti-skyrmion). In the following parts we call
such composite object "\emph{composite electron (hole)}".

Due to the Chern-Simon term $\frac{1}{4\pi}\epsilon^{\mu \nu \lambda}a_{\mu
}\partial_{\nu}a_{\lambda}$, the instantons are also confined by linear
potential and irrelevant to low energy physics. Thus the spinons are also
deconfined. FIG.17 shows the mass gap of $\mathbf{z}$ particle for the
parameter $t^{\prime}=0.033t$, $2m_{z}=4\pi c(\frac{1}{g_{c}}-\frac{1}{g})$.
One can see that $m_{z}$ is always much smaller than the mass gap of
electrons, $\Delta E$ as $m_{z}\ll \Delta E$. So the low energy physics is
dominated by $\mathbf{z}$ particle, the so-called composite electron (hole).

Secondly, we study the properties of gauge fluctuations. After integrating the
massive $\mathbf{z}$ particle, the effective Lagrangian for gauge field
$a_{\mu}$ becomes
\begin{align}
\mathcal{L}_{\mathrm{eff}}  &  =\frac{1}{4e_{a}^{2}}(\partial_{\mu}a_{\nu
})^{2}+\frac{1}{4\pi}\epsilon^{\mu \nu \lambda}a_{\mu}\partial_{\nu}a_{\lambda
}\nonumber \\
&  +\frac{1}{2\pi}\epsilon^{\mu \nu \lambda}A_{\mu}\partial_{\nu}a_{\lambda
}+\frac{1}{4\pi}\epsilon^{\mu \nu \lambda}A_{\mu}\partial_{\nu}A_{\lambda}.
\end{align}
Then the partition function of the effective model is written as
\[
\mathcal{Z}=\int \mathcal{D}\left[  a\right]  e^{-\int_{0}^{\beta}%
d\tau \mathcal{L}_{\mathrm{eff}}}.
\]
Then we introduce $a_{+,\mu}=A_{\mu}+a_{\mu},\ a_{-,\mu}=A_{\mu}-a_{\mu}\ $and
get the partition function as
\[
\mathcal{Z}=\int \mathcal{D}\left[  a_{+}\right]  e^{-\int_{0}^{\beta}%
d\tau \mathcal{L}_{\mathrm{eff}}},
\]
where
\begin{equation}
\mathcal{L}_{\mathrm{eff}}=\frac{1}{4e_{a}^{2}}(\partial_{\mu}a_{\nu,+}%
)^{2}+\frac{1}{4\pi}\epsilon^{\mu \nu \lambda}a_{\mu,+}\partial_{\nu}%
a_{\lambda,+}.
\end{equation}

With the Chern-Simons term $\frac{1}{4\pi}\epsilon^{\mu \nu \lambda}a_{\mu
,+}\partial_{\nu}a_{\lambda,+}$, the gauge field $a_{\mu,+}$ indicates
quantized spin-charge synchronized edge states and quantized spin-charge
synchronized Hall effect pointed out in Ref.\cite{he2}. The edge excitation is
described by the following one dimension fermion theory\cite{edge0,edge}
\begin{equation}
\mathcal{L}_{\text{\textrm{edge}}}=\tilde{\psi}^{\dag}(\partial_{t}-\tilde
{v}\partial_{x})\tilde{\psi}%
\end{equation}
where $\tilde{\psi}$ carries a unit of $a_{+,\mu}$ "charge". One can see a
chiral edge mode (the lines in FIG.16). Consequently, we get the spin-charge
synchronized Hall conductivity as
\begin{equation}
\tilde{\sigma}=\lim_{\omega \rightarrow0}{\frac{1}{\omega}}\epsilon
_{ij}\left \langle \tilde{J}_{i}(\omega,0)\tilde{J}_{j}(-\omega,0)\right \rangle
=\frac{e^{2}}{h}%
\end{equation}
where
\begin{equation}
\tilde{J}_{i}=i\langle \sum_{a}\bar{\psi}_{a}\gamma_{i}(1-\mathbf{n\cdot
\sigma)}\psi_{a}/2\rangle.
\end{equation}

\begin{figure}[ptb]
\includegraphics[width=0.5\textwidth]{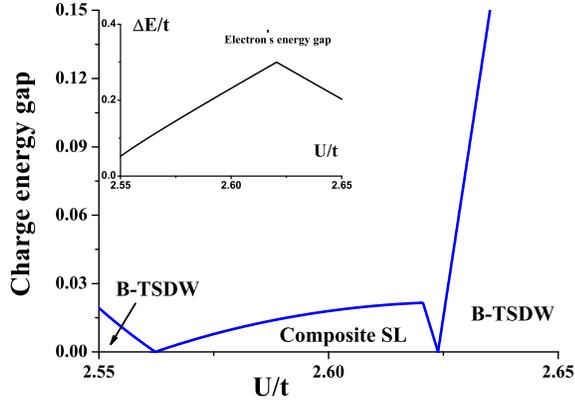}\caption{(Color online) The
charge energy gap for case of $t^{\prime}=0.033t$, $\varepsilon=0.15t$: the
charge carrier is composite electron. In composite SL, the charge energy gap
is that of spin gap, $m_{s}$; in B-TSDW, the charge energy gap is that of a
pair of skyrmion and anti-skyrmion, $\Delta_{c}$. The energy gap of fermion
quasi-particles are very big (see inset).}%
\end{figure}

Finally we use the duality relationship between spinons and skyrmions to learn
the quantum phase transition at $g=g_{c}$ dividing long range B-TSDW and short
range B-TSDW (composite SL).

In B-TSDW, we can define the skyrmion (or anti-skyrmion) with winding number
$\mathcal{Q}=\int \mathbf{d^{2}\mathbf{r}}\frac{1}{4\pi}\epsilon_{0\nu \lambda
}\mathbf{n_{\mathrm{s}}\cdot \partial}^{\nu}\mathbf{n_{\mathrm{s}}%
\times \partial}^{\lambda}\mathbf{n_{\mathrm{s}}=\pm}1,$ of which the solutions
in the continuum limit are \cite{pol}%
\begin{align}
\mathbf{n}_{\mathrm{s}}  &  =(\frac{\lambda(x-x_{0})}{|\mathbf{r}%
-\mathbf{r}_{0}|^{2}+\lambda^{2}},\text{ }\pm \frac{\lambda(y-y_{0}%
)}{|\mathbf{r}-\mathbf{r}_{0}|^{2}+\lambda^{2}},\text{ }\\
&  \pm \frac{\lambda}{|\mathbf{r}-\mathbf{r}_{0}|^{2}+\lambda^{2}}).\nonumber
\end{align}
Here $\lambda$ is the radius of the skyrmion at $\mathbf{r}_{0}=(x_{0},y_{0}%
)$. In long range B-TSDW, due to "spin-charge synchronized charge-flux
binding" effect, $Q=\pm1$ skyrmion carries a unit electric charge $q=\mp1$ and
a unit "charge" $q_{s}=\mp1$. With a unit "charge" $q_{s}$, a $Q=\pm1$
skyrmion gets half spin and becomes a charged $S=1/2$ fermion.

The mass of the skyrmion (or anti-skyrmion) is associated with
\[
m_{\mathrm{skyrmion}}=m_{\mathrm{anti-skyrmion}}=\frac{\rho_{\mathrm{s,eff}}%
}{2}\int{d^{2}\mathbf{{r}}}\left(  {\nabla \mathbf{{n}}}_{\mathrm{s}}\right)
^{2}=4\pi \rho_{\mathrm{s,eff}}%
\]
where $\rho_{\mathrm{s,eff}}=(1-\frac{g}{g_{c}})\rho_{\mathrm{s}}$. This
result indicates the charge gap is really the mass gap of a pair of
skyrmion-anti-skyrmion
\begin{equation}
\Delta_{c}=m_{\mathrm{skyrmion}}+m_{\mathrm{anti-skyrmion}}=8\pi(1-\frac
{g}{g_{c}})\rho_{\mathrm{s}}%
\end{equation}
that will close at the critical point $g=g_{c}$, $\Delta_{c}\rightarrow0$.
From FIG.5, one can see that there exists two QCPs ($g=g_{c}$) between B-TSDW
and composite SL, at which the charged excitations have no energy gap, while
the usual electrons without trapping spin texture still has big mass gap
$\Delta E\gg \Delta_{c}$ (see inset of FIG.17). In B-TSDW, the low energy
charge dynamics is dominated by fermionic charged skyrmions rather than the
electrons. At these QCPs, the system is a semi-metal with gapless charge
excitations. The dotted line in FIG.15 is the energy scale of the charge gap.

Finally we find that there exists a new type of spin liquid - composite spin
liquid. The low energy excitations are "\emph{composite electrons}" that are
$S=1/2$ charge $\pm e$ fermions with trapping a topological spin texture. See
illustration in FIG.18. At the QCPs between long range B-TSDW and composite
SL, the system becomes a semi-metal with gapless charge excitations (even for
gapped electrons).

\begin{figure}[ptb]
\includegraphics[width=0.5\textwidth]{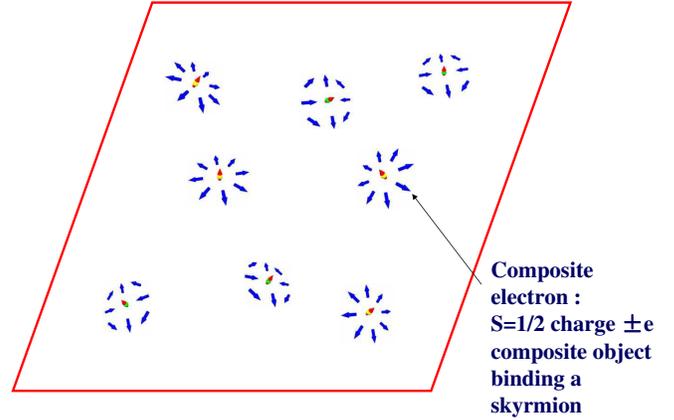}\caption{(Color online) The
illustration of composite spin liquid, of which the elementary excitations are
the "\emph{composite electrons}" that are $S=1/2$ charge $\pm e$ fermions with
trapping a topological spin texture.}%
\end{figure}

\begin{figure}[ptb]
\includegraphics[width=0.5\textwidth]{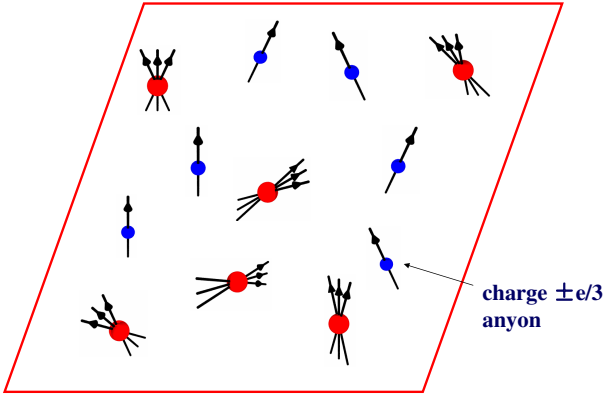}\caption{(Color online) The
illustration of Fractional quantum Hall state. Small blue balls with single
arrow denotes the anyonic excitations with $\pm e/3$ charge.}%
\end{figure}

\begin{figure}[ptb]
\includegraphics[width=0.5\textwidth]{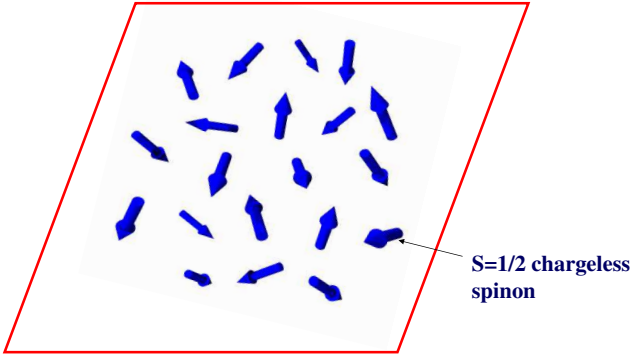}\caption{(Color online) The
illustration of quantum spin liquid. The blue arrows denote an $S=1/2$
chargeless spinons.}%
\end{figure}

\section{Conclusion and discussion}

In the end, we give a summary. We found a new type of topological state which
we name as composite spin liquid. Composite spin liquid state can be regarded
as a short range B-type topological spin-density-wave which is beyond the
classification of traditional spin liquid states. For traditional spin liquid
states, there always exists spin-charge separation. While for composite spin
liquid there is no spin-charge separation. Instead, the elementary excitations
are "composite electrons" with both spin degree of freedom and charge degree
of freedom, together with topological spin texture. This topological state
supports single chiral edge mode but no topological degeneracy. In addition,
the QCPs between long range B-TSDW and composite SL are also nontrivial, at
which the system becomes a semi-metal with gapless charge excitations (even
for gapped electrons).

In addition, we give a comparison on different exotic quantum orders beyond
Landau's theory:

\begin{widetext}
\begin{table*}[t]%
\begin{tabular}
[c]{|c|cccc|}\hline & Excitations & Topological degeneracy & Edge
state & Classification\\ \hline
FQH state & Charged anyon & Yes & Yes & K-matrix\\
TBI & Electron & No & Yes & Ten-fold way\\
SL & Spinon & -- & -- & PSG\\
Composite SL & Composite electron & No &
Yes & ?\\ \hline
\end{tabular}
\caption{The differences (the types of excitations, if there exists
topological degeneracy for the ground states on torus, if there
exist edge states, the way to classify the topological states)
between four exotic quantum orders beyond Landau's theory:
fractional quantum Hall state(FQH), topological band insulator
(TBI), spin liquid (SL), composite spin liquid (SL).}
\end{table*}
\end{widetext}

\begin{enumerate}
\item \textit{Fractional quantum Hall (FQH) state }: due to charge-flux
binding effect, the elementary excitations are anyonic excitations with
fractional electric charge and fractional quantized Hall
conductivity\cite{TSG8259,Laughlinc}. FIG.19 shows the anyonic excitations
with $\pm e/3$ charge (small blue balls with single arrow). The ground state
on a torus has topological degeneracy. For the open system, there exist chiral
edge states on its boundary. By their effective CS theories (or K-matrix
theory\cite{Kmat,read,fr}), people can classify fractional quantized Hall
states into different Abelian states and nonAbelian states;

\item \textit{Topological band insulator (TBI) : }the elementary excitations
are gapped electrons (or holes). For this topological state, there exist
gapless edge states. However, there is no topological degeneracy for the
ground state. By "ten-fold way" of random matrix, people classify topological
band insulators into $Z_{2}$ type or $Z$ type\cite{zi,ki,ry};

\item \textit{Spin liquid (SL) }: due to the big electron gap (Mott gap), the
excitations are deconfined spinons with only spin degree of freedom. In
FIG.20, the blue arrow denotes an $S=1/2$ chargeless spinon. By PSGs, people
classify quantum spin liquid states into $SU(2)$ type, $U(1)$ type or $Z_{2}$
type\cite{wen,wxg}. For topological spin liquid (for example chiral spin
liquid), there exist gapless edge states and topological degeneracy; while for
gapless spin liquid (for example algebraic spin liquid), there is no well
defined gapless edge states and topological degeneracy. For this reason we use
"$-$" to denote the uncertainty in table.1;

\item \textit{Composite spin liquid }: the elementary excitations are the
"composite electrons" with both spin degree of freedom and charge degree of
freedom, together with topological spin texture (See FIG.18). For this
topological states, there exist gapless edge states but no topological
degeneracy. Till now we don't know how to characterize composite spin liquid
states. For this reason we use "$?$" to denote the situation in table.1.
\end{enumerate}

Finally, we address the relevant experimental realization. This topological
Hubbard model on honeycomb lattice may be simulated in optical lattice of cold
atoms. In Ref.\cite{zhu}, it is proposed that the (spinless) Haldane model on
honeycomb optical lattice can be realized in the cold atoms. When
two-component fermions with repulsive interaction are put into such optical
lattice, one can get an effective topological Hubbard model. It is easy to
change the potential barrier by varying the laser intensities to tune the
Hamiltonian parameters including the hopping strength ($t$-term), the
staggered potential ($\varepsilon$-term) and the particle interaction ($U$-term).

\begin{acknowledgments}
This work is supported by NFSC Grant No. 10874017, 11174035, National Basic
Research Program of China (973 Program) under the grant No. 2011CB921803,
2012CB921704, 2011cba00102.
\end{acknowledgments}

\section{Appendix A: Theory of spin fluctuations - to get the O(3) nonlinear
$\sigma$ model}

The Hamiltonian of the topological Hubbard model on honeycomb lattice is given
by
\begin{align}
H  &  =-t\sum \limits_{\left \langle {i,j}\right \rangle ,\sigma}\left(  \hat
{c}_{i\sigma}^{\dagger}\hat{c}_{j\sigma}+h.c.\right)  -t^{\prime}%
\sum \limits_{\left \langle \left \langle {i,j}\right \rangle \right \rangle
,\sigma}e^{i\phi_{ij}}\hat{c}_{i\sigma}^{\dagger}\hat{c}_{j\sigma}\nonumber \\
&  +U\sum \limits_{i}\hat{n}_{i\uparrow}\hat{n}_{i\downarrow}+\mu
\sum \limits_{i,\sigma}\hat{c}_{i\sigma}^{\dagger}\hat{c}_{i\sigma}\nonumber \\
&  +\varepsilon \sum \limits_{i\in{A,}\sigma}\hat{c}_{i\sigma}^{\dagger}\hat
{c}_{i\sigma}-\varepsilon \sum \limits_{i\in{B,}\sigma}\hat{c}_{i\sigma
}^{\dagger}\hat{c}_{i\sigma}.
\end{align}
$t$ and $t^{\prime}$ are the nearest neighbor and the next nearest neighbor
hoppings, respectively.\ We introduce a complex phase $\phi_{ij}$ $\left(
\left \vert \phi_{ij}\right \vert =\frac{\pi}{2}\right)  $ to the next nearest
neighbor hopping, of which the positive phase is set to be clockwise. $U$ is
the on-site Coulomb repulsion. $\sigma$ are the spin-indices representing
spin-up $(\sigma=\uparrow)$ and spin-down $(\sigma=\downarrow)$ for electrons,
$\varepsilon$ denotes an on-site staggered energy and is set to be $0.15t$.\ .

For free fermions (the on-site Coulomb repulsion $U$ is zero), the spectrum%
\begin{equation}
E_{\mathbf{k}}=\sqrt{\left \vert \xi_{k}\right \vert ^{2}+\left(  \xi
_{k}^{\prime}+\varepsilon \right)  ^{2}} \label{4}%
\end{equation}
where
\begin{equation}
\left \vert \xi_{\mathbf{k}}\right \vert =t\sqrt{3+2\cos{(\sqrt{3}k_{y})}%
+4\cos{(3k_{x}/2)}\cos{(\sqrt{3}k_{y}/2)}}\nonumber
\end{equation}
and
\begin{equation}
\xi_{k}^{\prime}=2t^{\prime}\sum \limits_{i}\sin{(\mathbf{k}\cdot \mathbf{b}%
_{i})}.
\end{equation}
Where ${\mathbf{b}_{i}}$\ is the nest nearest vectors. According to this
spectrum $E_{\mathbf{k}}$, we can see that there exist energy gaps
$\Delta_{f1}$, $\Delta_{f2}$ near points $\mathbf{k}_{1}=-\frac{2\pi}{3}(1,$
$1/\sqrt{3})$ and $\mathbf{k}_{2}=\frac{2\pi}{3}(1,$ $1/\sqrt{3})$ as
$\Delta_{f1}=\left \vert 2\varepsilon-6\sqrt{3}t^{\prime}\right \vert $ and
$\Delta_{f2}=2\varepsilon+6\sqrt{3}t^{\prime},$ respectively. There exist two
phases separated by the phase boundary $2\varepsilon=6\sqrt{3}t^{\prime}$, the
quantum anomalous Hall (QAH) state and the normal band insulator (BI) state
with trivial topological properties.

Because the Hubbard model on bipartite lattices is unstable against
antiferromagnetic instability, at half-filling, the ground state may be an
insulator with AF-SDW\ order with increasing interacting strength. Such AF-SDW
order is described by the following mean field order parameter $\langle
(-1)^{i}\hat{c}_{i}^{\dag}\sigma^{z}\hat{c}_{i}\rangle=M.$ Here $M$ is the
staggered magnetization. In the mean field theory, the Hamiltonian of the
topological Hubbard model is obtained as%
\begin{equation}
H_{MF}=H-\sum \limits_{i}(-1)^{i}\Delta_{M}\hat{c}_{i}^{\dagger}\sigma_{z}%
\hat{c}_{i}%
\end{equation}
where $\Delta_{M}=\frac{UM}{2}$. Then in the momentum space we get
\begin{equation}
H=\sum_{k}c_{k}^{\dagger}h_{k}c_{k},
\end{equation}
where $c_{k}^{\dagger}=(c_{k,A\uparrow}^{\dagger},c_{k,A\downarrow}^{\dagger
},c_{k,B\uparrow}^{\dagger},c_{k,B\downarrow}^{\dagger})$ and
\[
h_{k}=\left(
\begin{array}
[c]{cc}%
\xi_{\mathbf{k}^{\prime}}+\frac{UM}{2}\sigma_{z}+\varepsilon & \xi
_{\mathbf{k}}\\
\left(  \xi_{\mathbf{k}}\right)  ^{\ast} & -\xi_{\mathbf{k}^{\prime}}%
-\frac{UM}{2}\sigma_{z}-\varepsilon
\end{array}
\right)  .
\]
After diagonalization, we can get the quasi-particles spectrums
\begin{equation}
E_{\mathbf{k}_{1}}=\pm \sqrt{(\xi_{k}^{\prime}+\Delta_{M}+\varepsilon)^{2}%
+|\xi_{k}|^{2}}%
\end{equation}
and
\begin{equation}
E_{\mathbf{k}_{2}}=\pm \sqrt{(\xi_{k}^{\prime}-\Delta_{M}+\varepsilon)^{2}%
+|\xi_{k}|^{2}}.
\end{equation}

By minimizing the ground state's energy, the self-consistent equation in the
reduced BZ is reduced into
\begin{equation}
1=\frac{1}{N_{s}M}\sum \limits_{\mathbf{k\in}BZ}{[\frac{\xi_{\mathbf{k}%
}^{\prime}+\Delta_{M}+\varepsilon}{2E_{\mathbf{k}_{1}}}-\frac{\xi_{\mathbf{k}%
}^{\prime}-\Delta_{M}+\varepsilon}{2E_{\mathbf{k}_{2}}}{]}}%
\end{equation}
where $N_{s}$ is the number of unit cells. The phase diagram has been obtained
in Ref.\cite{he2}. There are totally five phases, NI state, QAH state, A-TSDW
state, B-TSDW state and trivial AF-SDW state seperated by two types of phase
transitions : one is the magnetic phase transition [denoted by $(U/t)_{M}$]
between a magnetic order state with $M\neq0$ and a non-magnetic state with
$M=0$, the other one is the topological quantum phase transition [denoted by
$(\frac{U}{t})_{c1}$ or $(\frac{U}{t})_{c2}$] that is characterized by the
condition of zero fermion's energy gaps, $\Delta_{f1}=-6\sqrt{3}t^{\prime
}+2\varepsilon+UM=0$ or $\Delta_{f2}=6\sqrt{3}t^{\prime}+2\varepsilon-UM=0$.

We deal with the spin fluctuations by using the path-integral formulation of
electrons with spin rotation symmetry. The interaction term can be handled by
using the SU(2) invariant Hubbard-Stratonovich decomposition in the arbitrary
on-site unit vector $\mathbf{\Omega}_{i}$
\begin{equation}
\hat{n}_{i\uparrow}\hat{n}_{i\downarrow}=\frac{\left(  \hat{c}_{i}^{\dagger
}\hat{c}_{i}\right)  ^{2}}{4}-\frac{1}{4}[\mathbf{\Omega}_{i}\mathbf{\cdot
}\hat{c}_{i}^{\dag}\mathbf{\sigma}\hat{c}_{i}]^{2}.
\end{equation}
Here $\mathbf{\sigma=}\left(  \sigma_{x},\sigma_{y},\sigma_{z}\right)  $ are
the Pauli matrices. By replacing the electronic operators $\hat{c}%
_{i}^{\dagger}$ and $\hat{c}_{j}$ by Grassmann variables $c_{i}^{\ast}$ and
$c_{j}$, the effective Lagrangian of the 2D generalized Hubbard model at half
filling is obtained:%
\begin{align}
\mathcal{L}_{\mathrm{eff}}  &  =\sum_{i,\sigma}c_{i\sigma}^{\ast}%
\partial_{\tau}c_{i,\sigma}-t\sum \limits_{\left \langle {i,j}\right \rangle
,\sigma}\left(  c_{i\sigma}^{\ast}c_{j\sigma}+h.c.\right) \nonumber \\
&  -t^{\prime}\sum \limits_{\left \langle \left \langle {i,j}\right \rangle
\right \rangle ,\sigma}e^{i\phi_{ij}}c_{i\sigma}^{\ast}c_{j\sigma}-\Delta
_{M}\sum_{i{,\sigma}}c_{i{,\sigma}}^{\ast}\mathbf{\Omega}_{i}\mathbf{\cdot
\sigma}c_{i{,\sigma}}\nonumber \\
&  +\varepsilon \sum \limits_{i\in{A,\sigma}}c_{i\sigma}^{\ast}c_{i\sigma
}-\varepsilon \sum \limits_{i\in{B,\sigma}}c_{i\sigma}^{\ast}c_{i\sigma}.
\end{align}

To describe the spin fluctuations, we use the Haldane's mapping:
\begin{equation}
\mathbf{\Omega}_{i}=(-1)^{i}\mathbf{n}_{i}\sqrt{1-\mathbf{L}_{i}^{2}%
}+\mathbf{L}_{i}%
\end{equation}
where $\mathbf{n}_{i}=(n_{i}^{x},n_{i}^{y},n_{i}^{z})$ is the Neel vector that
corresponds to the long-wavelength part of $\mathbf{\Omega}_{i}$ with a
restriction $\mathbf{n}_{i}^{2}=1.$ $\mathbf{L}_{i}$ is the transverse canting
field that corresponds to the short-wavelength parts of $\mathbf{\Omega}_{i}$
with a restriction $\mathbf{L}_{i}\cdot \mathbf{n}_{i}=0$. We then rotate
$\mathbf{\Omega}_{i}$ to $\mathbf{\hat{z}}$-axis for the spin indices of the
electrons at $i$-site:%
\begin{align}
\psi_{i}  &  =U_{i}^{\dagger}c_{i}\\
U_{i}^{\dagger}\mathbf{n}_{i}\cdot \mathbf{\sigma}U_{i}  &  =\mathbf{\sigma
}_{z}\\
U_{i}^{\dagger}\mathbf{L}_{i}\cdot \mathbf{\sigma}U_{i}  &  =\mathbf{l}%
_{i}\cdot \mathbf{\sigma}%
\end{align}
where $U_{i}\in$\textrm{SU(2)/U(1)}.

One then can derive the following effective Lagrangian after such spin
transformation:%
\begin{align}
\mathcal{L}_{\mathrm{eff}}  &  =\sum \limits_{i,\sigma}\psi_{i,\sigma}^{\ast
}\partial_{\tau}\psi_{i,\sigma}+\sum_{i,\sigma}\psi_{i,\sigma}^{\ast}%
a_{0}\left(  i\right)  \psi_{i,\sigma}\nonumber \\
&  -t\sum \limits_{\left \langle {i,j}\right \rangle ,\sigma}\left(
\psi_{i,\sigma}^{\ast}e^{ia_{ij}}\psi_{j,\sigma}+h.c.\right)  -t^{\prime}%
\sum \limits_{\left \langle \left \langle {i,j}\right \rangle \right \rangle
,\sigma}e^{i\phi_{ij}}\psi_{i,\sigma}^{\ast}e^{ia_{ij}}\psi_{j,\sigma
}\nonumber \\
&  +\varepsilon \sum \limits_{i\in{A,\sigma}}\psi_{i,\sigma}^{\ast}e^{ia_{ii}%
}\psi_{i,\sigma}-\varepsilon \sum \limits_{i\in{B,\sigma}}\psi_{i,\sigma}^{\ast
}e^{ia_{ii}}\psi_{i,\sigma}\nonumber \\
&  -\Delta_{M}\sum_{i{,\sigma}}\psi_{i,\sigma}^{\ast}\left[  \left(
-1\right)  ^{i}\mathbf{\sigma}_{z}\sqrt{1-\mathbf{l}_{i}^{2}}+\mathbf{l}%
_{i}\cdot \mathbf{\sigma}\right]  \psi_{i,\sigma}%
\end{align}
where the auxiliary gauge fields $a_{ij}=a_{ij,1}\sigma_{x}+a_{ij,2}\sigma
_{y}$ and $a_{0}\left(  i\right)  =a_{0,1}\left(  i\right)  \sigma_{x}%
+a_{0,2}\left(  i\right)  \sigma_{y}\ $are defined as
\begin{equation}
e^{ia_{ij}}=U_{i}^{\dagger}U_{j},\text{ }a_{0}\left(  i\right)  =U_{i}%
^{\dagger}\partial_{\tau}U_{i}. \label{ao}%
\end{equation}

In terms of the mean field result $M=\left(  -1\right)  ^{i}\langle \psi
_{i}^{\ast}\mathbf{\sigma}_{z}\psi_{i}\rangle$ as well as the approximations,
\begin{equation}
\sqrt{1-\mathbf{l}_{i}^{2}}\simeq1-\frac{\mathbf{l}_{i}^{2}}{2},\text{
}e^{ia_{ij}}\simeq1+ia_{ij},\nonumber
\end{equation}
we obtain the effective Hamiltonian as:%

\begin{align}
\mathcal{L}_{\mathrm{eff}}  &  \simeq \sum \limits_{i,\sigma}\psi_{i,\sigma
}^{\ast}\partial_{\tau}\psi_{i,\sigma}+\sum_{i,\sigma}\psi_{i,\sigma}^{\ast
}\left(  a_{0}\left(  i\right)  -\Delta \mathbf{l}_{i}\cdot \mathbf{\sigma
}\right)  \psi_{i,\sigma}\nonumber \\
&  -\Delta_{M}\sum_{i{,\sigma}}\left(  -1\right)  ^{i}\psi_{i,\sigma}^{\ast
}\mathbf{\sigma}_{z}\psi_{i,\sigma}-t\sum \limits_{\left \langle {i,j}%
\right \rangle ,\sigma}\psi_{i,\sigma}^{\ast}\left(  1+ia_{ij}\right)
\psi_{j,\sigma}\nonumber \\
&  -t^{\prime}\sum \limits_{\left \langle \left \langle {i,j}\right \rangle
\right \rangle ,\sigma}e^{i\phi_{ij}}\psi_{i,\sigma}^{\ast}\left(
1+ia_{ij}\right)  \psi_{j,\sigma}+\Delta M\sum_{i{,\sigma}}\frac
{\mathbf{l}_{i}^{2}}{2}\nonumber \\
&  +\varepsilon \sum \limits_{i\in{A,\sigma}}\psi_{i,\sigma}^{\ast}\left(
1+ia_{ii}\right)  \psi_{i,\sigma}-\varepsilon \sum \limits_{i\in{B,\sigma}}%
\psi_{i,\sigma}^{\ast}\left(  1+ia_{ii}\right)  \psi_{i,\sigma}%
\end{align}
By integrating out the fermion fields $\psi_{i}^{\ast}$ and $\psi_{i},$ the
effective action with the quadric terms of $[a_{0}\left(  i\right)
-\Delta \mathbf{\sigma \cdot l}_{i}]$ and $a_{ij}$ becomes
\begin{equation}
\mathcal{S}_{\mathrm{eff}}=\frac{1}{2}\int_{0}^{\beta}d\tau \sum_{i}%
[-4\varsigma(a_{0}\left(  i\right)  -\Delta_{M}\mathbf{\sigma \cdot l}_{i}%
)^{2}+4\rho_{s}a_{ij}^{2}+\frac{2\Delta_{M}^{2}}{U}\mathbf{l}_{i}^{2}].
\label{eff1}%
\end{equation}

To give $\rho_{s}$ and $\varsigma$ for calculation in detail, we choose
$U_{i}$ to be%
\begin{equation}
U_{i}=(%
\begin{array}
[c]{cc}%
z_{i\uparrow}^{\ast} & z_{i\downarrow}^{\ast}\\
-z_{i\downarrow} & z_{i\uparrow}%
\end{array}
),
\end{equation}
where $\mathbf{n}_{i}=\mathbf{\bar{z}}_{i}\mathbf{\sigma z}_{i},$
$\mathbf{z}_{i}=\left(  z_{i\uparrow},z_{i\downarrow}\right)  ^{T},$
$\mathbf{\bar{z}}_{i}\mathbf{z}_{i}\mathbf{=1.}$ And the spin fluctuations
around $\mathbf{n}_{i}=\mathbf{\hat{z}}_{i}$ is
\begin{align}
\mathbf{n}_{i}  &  =\mathbf{\hat{z}}_{i}\mathbf{+}\text{Re}\left(
\mathbf{\phi}_{i}\right)  \mathbf{\hat{x}+}\text{Im}\left(  \mathbf{\phi}%
_{i}\right)  \mathbf{\hat{y}}\\
\mathbf{z}_{i}  &  =\left(
\begin{array}
[c]{c}%
1-\left \vert \mathbf{\phi}_{i}\right \vert ^{2}/8\\
\mathbf{\phi}_{i}/2
\end{array}
\right)  +O(\mathbf{\phi}_{i}^{3}).
\end{align}
Then the quantities $U_{i}^{\dagger}U_{j}$ and $U_{i}^{\dagger}\partial_{\tau
}U_{i}$ can be expanded in the power of $\mathbf{\phi}_{i}-\mathbf{\phi}_{j}$
and $\partial_{\tau}\mathbf{\phi}_{i},$%
\begin{align}
U_{i}^{\dagger}U_{j}  &  =e^{-i\frac{\mathbf{\phi}_{i}-\mathbf{\phi}_{j}}%
{2}\sigma_{y}}\\
U_{i}^{\dagger}\partial_{\tau}U_{i}  &  =\left(
\begin{array}
[c]{cc}%
0 & \frac{1}{2}\partial_{\tau}\mathbf{\phi}_{i}\\
-\frac{1}{2}\partial_{\tau}\mathbf{\phi}_{i} & 0
\end{array}
\right)  .
\end{align}
According to Eq.(\ref{ao}), the gauge field $a_{ij}$ and $a_{0}\left(
i\right)  $ are given as
\begin{align}
a_{ij}  &  =-\frac{1}{2}\left(  \mathbf{\phi}_{i}-\mathbf{\phi}_{j}\right)
\mathbf{\sigma}_{y}\\
a_{0}\left(  i\right)   &  =\frac{i}{2}\partial_{\tau}\mathbf{\phi}%
_{i}\mathbf{\sigma}_{y}.
\end{align}
Supposing $a_{ij}$ and $a_{0}\left(  i\right)  $ to be a constant in space and
denoting $\mathbf{\partial}_{i}\mathbf{\mathbf{\phi}}_{i}\mathbf{=a}$ and
$\partial_{\tau}\mathbf{\phi}_{i}=iB_{y}$, we have
\begin{align}
a_{ij}  &  =-\frac{1}{2}\mathbf{a\cdot(i-j)\sigma}_{y}\\
a_{0}\left(  i\right)   &  =-\frac{1}{2}B_{y}\mathbf{\sigma}_{y}.
\end{align}
The energy of Hamiltonian of Eq.(\ref{eff1}) becomes
\begin{equation}
E\left(  B_{y},\mathbf{a}\right)  =-\frac{1}{2}\zeta B_{y}^{2}+\frac{1}{2}%
\rho_{s}\mathbf{a}^{2}. \label{ener}%
\end{equation}
Then one could get $\zeta$ and $\rho_{s}$ from the following equations by
calculating the partial derivative of the energy
\begin{align}
\zeta &  =-\frac{1}{N}\frac{\partial^{2}E_{0}\left(  B_{y}\right)  }{\partial
B_{y}^{2}}|_{B_{y}=0}\\
\rho_{s}  &  =\frac{1}{N}\frac{\partial^{2}E_{0}\left(  \mathbf{a}\right)
}{\partial \mathbf{a}^{2}}|_{\mathbf{a}=0}.
\end{align}
Here $E_{0}\left(  B_{y}\right)  $ and $E_{0}\left(  \mathbf{a}\right)  $ are
the energy of the lower Hubbard band%
\begin{align}
E_{0}\left(  B_{y}\right)   &  =\sum \limits_{\mathbf{k}}\left(
E_{+,\mathbf{k}}^{\zeta}+E_{-,\mathbf{k}}^{\zeta}\right) \label{ek}\\
E_{0}\left(  \mathbf{a}\right)   &  =\sum \limits_{\mathbf{k}}\left(
E_{+,\mathbf{k}}^{\rho}+E_{-,\mathbf{k}}^{\rho}\right)  \label{er}%
\end{align}
where $E_{+,\mathbf{k}}^{\zeta},$ $E_{-,\mathbf{k}}^{\zeta}$ and
$E_{+,\mathbf{k}}^{\rho},$ $E_{-,\mathbf{k}}^{\rho}$ are the energies of the
following Hamiltonian $\mathcal{H}^{\zeta}$ and $\mathcal{H}^{\rho}$%
\begin{align}
\mathcal{H}^{\zeta}  &  =-t\sum \limits_{\left \langle {i,j}\right \rangle
,\sigma}\left(  \psi_{i,\sigma}^{\ast}\psi_{j,\sigma}+h.c.\right)  -t^{\prime
}\sum \limits_{\left \langle \left \langle {i,j}\right \rangle \right \rangle
,\sigma}e^{i\phi_{ij}}\psi_{i,\sigma}^{\ast}\psi_{j,\sigma}\nonumber \\
&  +\varepsilon \sum \limits_{i\in{A,\sigma}}\psi_{i,\sigma}^{\ast}%
\psi_{i,\sigma}-\varepsilon \sum \limits_{i\in{B,\sigma}}\psi_{i,\sigma}^{\ast
}\psi_{i,\sigma}\nonumber \\
&  +\sum_{i,\sigma}\psi_{i,\sigma}^{\ast}a_{0}\left(  i\right)  \psi
_{i,\sigma}-\Delta_{M}\sum_{i{,\sigma}}\left(  -1\right)  ^{i}\psi_{i,\sigma
}^{\ast}\mathbf{\sigma}_{z}\psi_{i,\sigma},
\end{align}%
\begin{align}
\mathcal{H}^{\rho}  &  =-t\sum \limits_{\left \langle {i,j}\right \rangle
,\sigma}\psi_{i,\sigma}^{\ast}e^{ia_{ij}}\psi_{j,\sigma}-t^{\prime}%
\sum \limits_{\left \langle \left \langle {i,j}\right \rangle \right \rangle
,\sigma}e^{i\phi_{ij}}\psi_{i,\sigma}^{\ast}e^{ia_{ij}}\psi_{j,\sigma
}\nonumber \\
&  +\varepsilon \sum \limits_{i\in{A,\sigma}}\psi_{i,\sigma}^{\ast}e^{ia_{ii}%
}\psi_{i,\sigma}-\varepsilon \sum \limits_{i\in{B,\sigma}}\psi_{i,\sigma}^{\ast
}e^{ia_{ii}}\psi_{i,\sigma}\nonumber \\
&  -\Delta_{M}\sum_{i{,\sigma}}\left(  -1\right)  ^{i}\psi_{i,\sigma}^{\ast
}\mathbf{\sigma}_{z}\psi_{i,\sigma}%
\end{align}

Using the Fourier transformation for $\mathcal{H}^{\zeta}$, we have the
spectrum of the lower band of $\mathcal{H}^{\zeta}$: \begin{widetext}
\begin{equation}
E_{\pm,\mathbf{k}}^{^{\zeta}}=-\frac{1}{2}\sqrt{4|\xi_{k}|^{2}+2a^{2}%
+B_{y}^{2}+2d^{2}\pm2\sqrt{a^{4}+B_{y}^{2}a^{2}-2a^{2}d^{2}+B_{y}^{2}%
d^{2}+4B_{y}^{2}|\xi_{k}|^{2}+d^{4}+2adB_{y}^{2}}} \label{es}%
\end{equation}
\end{widetext}where $a=\xi_{k^{\prime}}+\frac{UM}{2}+\varepsilon$ and
$d=\xi_{k^{\prime}}-\frac{UM}{2}+\varepsilon$.

Using $\varsigma=-\frac{1}{N}\frac{\partial^{2}E_{0}{(B_{y})}}{\partial
B_{y}^{2}}|_{B_{y}=0}$ and $E_{0}{(B_{y})}=\sum \limits_{\mathbf{k}}\left(
E_{+,\mathbf{k}}^{\varsigma}+E_{-,\mathbf{k}}^{\varsigma}\right)  $, we can
get $\varsigma$ to be%
\begin{align}
\varsigma &  =\frac{-1}{N_{s}}\sum \limits_{\mathbf{k}}\frac{1}{8\sqrt{2}%
}(\frac{-2+\frac{2\left[  4|\xi_{k}|^{2}+(a+d)^{2}\right]  }{\sqrt
{(a^{2}-d^{2})^{2}}}}{\sqrt{a^{2}+2|\xi_{k}|^{2}+d^{2}-\sqrt{(d^{2}-a^{2}%
)^{2}}}}\nonumber \\
&  -\frac{2+\frac{2\left[  4|\xi_{k}|^{2}+(a+d)^{2}\right]  }{\sqrt
{(a^{2}-d^{2})^{2}}}}{\sqrt{a^{2}+2|\xi_{k}|^{2}+d^{2}+\sqrt{(d^{2}-a^{2}%
)^{2}}}}).
\end{align}
Similarly, using the Fourier transformation for $\mathcal{H}^{\rho}$, we have
the spectrum of the lower band of $\mathcal{H}^{\rho}$: \begin{widetext}
\begin{align}
E_{\pm,\mathbf{k}}^{\rho}= &  -\frac{1}{2}\bigg(4|\psi|^{2}+2G^{2}%
-4B^{2}+4|\varphi|^{2}+2A^{2}\nonumber \\
&  \pm2\Big(4|\psi|^{2}G^{2}-8AG|\psi|^{2}+8AB\psi^{\ast}\varphi \\
&  -8AB\varphi^{\ast}\psi+8B\psi^{\ast}\varphi G-8\varphi^{\ast}B\psi
G-2A^{2}G^{2}-4(\varphi^{\ast}\psi-\psi^{\ast}\varphi)^{2}\nonumber \\
&  -8AGB^{2}-4G^{2}B^{2}+G^{4}-4B^{2}A^{2}+A^{4}+4A^{2}|\psi|^{2}%
\Big)^{\frac{1}{2}}\bigg)^{\frac{1}{2}}\label{eps}%
\end{align}
\end{widetext}where
\begin{align*}
A  &  =2t^{\prime}\sum \limits_{i}\cos{(\frac{1}{2}\mathbf{a}\cdot
\mathbf{b}_{i})}\sin{(\mathbf{k}\cdot \mathbf{b}_{i})}+\varepsilon+\frac{UM}%
{2},\\
G  &  =2t^{\prime}\sum \limits_{i}\cos{(\frac{1}{2}\mathbf{a}\cdot
\mathbf{b}_{i})}\sin{(\mathbf{k}\cdot \mathbf{b}_{i})}+\varepsilon-\frac{UM}%
{2},\\
B  &  =-2it^{\prime}\sum \limits_{i}\sin{(\frac{1}{2}\mathbf{a}\cdot
\mathbf{b}_{i})}\cos{(\mathbf{k}\cdot \mathbf{b}_{i}),}\\
\varphi &  =-t\sum \limits_{\delta}e^{i\mathbf{k}\cdot \delta}\cos{(\frac{1}%
{2}{\mathbf{a}}\cdot \delta),}\\
\psi &  =-t\sum \limits_{\delta}e^{i\mathbf{k}\cdot \delta}\sin({\frac{1}%
{2}{\mathbf{a}}\cdot \delta).}%
\end{align*}

Using $\rho_{s}=\frac{1}{N}\frac{\partial^{2}E_{0}(\mathbf{a})}{\partial
\mathbf{a}^{2}}|_{\mathbf{a}=0}$ and $E_{0}(\mathbf{a})=\sum
\limits_{\mathbf{k}}{\left(  E_{+,\mathbf{k}}^{\rho}+E_{-,\mathbf{k}}^{\rho
}\right)  }$, we can get $\rho_{s}=\rho_{s1}+\rho_{s2}$
where\begin{widetext}
\begin{align}
\rho_{s1} &  =\frac{-1}{N_{s}}\sum \limits_{\mathbf{k}}\bigg(-9t\cos{(\frac{3k_{x}}{2})}\cos{(\frac{\sqrt{3}k_{y}}{2}%
)}-36t^{\prime}(-1+\cos{(\sqrt{3}k_{y})}\sin^{2}{(\frac{3k_{x}}{2})}\nonumber \\
&  -9t^{\prime}(P+Q)\cos{(\frac{3k_{x}}{2})}\sin{(\frac{\sqrt{3}k_{y}}{2}%
)}+\Big(3(t^{2}((P-Q)^{2}+6t^{2})+6t^{\prime2}(P+Q)^{2}\nonumber \\
&  -6(t^{4}+t^{\prime2}(P+Q)^{2})\cos{(3k_{x})})-4t^{2}(P-Q)^{2}\cos
{(\frac{3k_{x}}{2})}\cos{(\frac{\sqrt{3}k_{y}}{2})}\nonumber \\
&  +(t^{2}((P-Q)^{2}+18t^{2})-18t^{\prime}(P+Q)^{2}\nonumber \\
&  +18(t^{\prime}(P+Q)^{2}-t^{4})\cos{(3k_{x})})\cos{(\sqrt{3}k_{y})}\nonumber \\
&  -9t^{\prime}(P-Q)^{2}(P+Q)\cos{(\frac{3k_{x}}{2})}\sin{(\frac{\sqrt{3}k_{y}}%
{2})}\nonumber \\
&  +72t^{2}t^{\prime}(P+Q)\sin^{2}{(\frac{3k_{x}}{2})}\sin{(\sqrt{3}k_{y}%
)}\Big)/\sqrt{(P^{2}-Q^{2})^{2}}\bigg)\nonumber \\
&  /8\sqrt{2}\sqrt{P^{2}+Q^{2}+2|\xi_{k}|^{2}+\sqrt{(P^{2}-Q^{2})^{2}}%
}\label{psjieguo1}%
\end{align}
\end{widetext}

and \begin{widetext}
\begin{align}
\rho_{s2}  &  =\frac{-1}{N_{s}}\sum \limits_{\mathbf{k}}\bigg(-9t\cos{(\frac{3k_{x}}{2})}\cos{(\frac{\sqrt{3}k_{y}}%
{2})}-36t^{\prime}(-1+\cos{(\sqrt{3}k_{y})}\sin^{2}{(\frac{3k_{x}}{2})}\nonumber \\
&  -9t^{\prime}(P+Q)\cos{(\frac{3k_{x}}{2})}\sin{(\frac{\sqrt{3}k_{y}}{2}%
)}-\Big(3(t^{2}((P-Q)^{2}+6t^{2})+6t^{\prime2}(P+Q)^{2}\nonumber \\
&  -6(t^{4}+t^{\prime2}(P+Q)^{2})\cos{(3k_{x})})-4t^{2}(P-Q)^{2}\cos
{(\frac{3k_{x}}{2})}\cos{(\frac{\sqrt{3}k_{y}}{2})}+(t^{2}((P-Q)^{2}%
+18t^{2})\nonumber \\
&
-18t^{\prime}(P+Q)^{2}+18(t^{\prime}(P+Q)^{2}-t^{4})\cos{(3k_{x})})\cos{(\sqrt
{3}k_{y})}\nonumber \\
&  -9t^{\prime}(P-Q)^{2}(P+Q)\cos{(\frac{3k_{x}}{2})}\sin{(\frac{\sqrt{3}k_{y}}%
{2})}+72t^{2}t^{\prime}(P+Q)\sin^{2}{(\frac{3k_{x}}{2})}\sin{(\sqrt{3}k_{y}%
)}\Big)/\sqrt{(P^{2}-Q^{2})^{2}}\bigg)\nonumber \\
&
/8\sqrt{2}\sqrt{P^{2}+Q^{2}+2|\xi_{k}|^{2}-\sqrt{(P^{2}-Q^{2})^{2}}}
\label{psjieguo2}%
\end{align}
\end{widetext}where $P=\xi_{k^{\prime}}-\frac{UM}{2}+\varepsilon,$
$Q=\xi_{k^{\prime}}+\frac{UM}{2}+\varepsilon.$

In addition, we study the continuum theory of the effective action. In the
continuum limit, we denote $\mathbf{n}_{i}$, $\mathbf{l}_{i}$, $ia_{ij}\simeq
U_{i}^{\dagger}U_{j}-1$ and $a_{0}\left(  i\right)  =U_{i}^{\dagger}%
\partial_{\tau}U_{i}$ by $\mathbf{n}(x,y)$, $\mathbf{l}(x,y)$, $U^{\dagger
}\partial_{x}U$ (or $U^{\dagger}\partial_{y}U$) and $U^{\dagger}\partial
_{\tau}U,$ respectively. From the relations between $U^{\dagger}\partial_{\mu
}U$ and $\partial_{\mu}\mathbf{n,}$
\begin{align}
a_{\tau}^{2}  &  =a_{\tau,1}^{2}+a_{\tau,2}^{2}=-\frac{1}{4}(\partial_{\tau
}\mathbf{n})^{2},\text{ }\tau=0,\label{tao}\\
a_{\mu}^{2}  &  =a_{\mu,1}^{2}+a_{\mu,2}^{2}=\frac{1}{4}(\partial_{\mu
}\mathbf{n})^{2},\text{ }\mu=x,y,\label{miu}\\
\mathbf{a}_{0}\mathbf{\cdot l}  &  \mathbf{=}\mathbf{-}\frac{i}{2}\left(
\mathbf{n}\times \partial_{\tau}\mathbf{n}\right)  \cdot \mathbf{l,} \label{dot}%
\end{align}
the continuum formulation of the action turns into
\begin{align}
\mathcal{S}_{\mathrm{eff}}  &  =\frac{1}{2}\int_{0}^{\beta}d\tau \int
d^{2}\mathbf{r}[\varsigma(\partial_{\tau}\mathbf{n)}^{2}+\rho_{s}\left(
\mathbf{\bigtriangledown n}\right)  ^{2}\nonumber \\
&  -4i\Delta_{M}\varsigma \left(  \mathbf{n}\times \partial_{\tau}%
\mathbf{n}\right)  \cdot \mathbf{l}+(\frac{2\Delta_{M}^{2}}{U}-4\Delta_{M}%
^{2}\varsigma)\mathbf{l}^{2}]
\end{align}
where the vector $\mathbf{a}_{0}$ is defined as $\mathbf{a}_{0}=\left(
a_{0,1},\text{ }a_{0,2},\text{ }0\right)  .$

Finally we integrate the transverse canting field $\mathbf{l}$ and obtain the
effective $\mathrm{NL}\sigma \mathrm{M}$ as
\begin{equation}
\mathcal{S}_{\mathrm{eff}}=\frac{1}{2g}\int_{0}^{\beta}d\tau \int d^{2}%
r[\frac{1}{c}\left(  \partial_{\tau}\mathbf{n}\right)  ^{2}+c\left(
\mathbf{\bigtriangledown n}\right)  ^{2}]\text{ } \label{non1}%
\end{equation}
with a constraint $\mathbf{n}^{2}=1.$ The coupling constant $g$ and spin wave
velocity $c$ are defined as: $g=\sqrt{\frac{1}{\rho_{s}\chi^{\perp}}},$
$c^{2}=\frac{\rho_{s}}{\chi^{\perp}}$ and $\chi^{\perp}\ $is the transverse
spin susceptibility%
\begin{equation}
\chi^{\perp}=[\varsigma^{-1}-2U]^{-1}.
\end{equation}

In addition, we need to determine another important parameter - the cutoff
$\Lambda$. On the one hand, the effective $\mathrm{NL}\sigma \mathrm{M}$ is
valid within the energy scale of electrons's gap, $\Delta E.$ On the other
hand, the lattice constant is a natural cutoff. Thus the cutoff is defined as
the following equation
\begin{equation}
\Lambda=\min(1,\frac{\Delta E}{c}).
\end{equation}

\section{Appendix B: Induced CSH terms}

In this appendix we will derive the low energy effective theory of (T-)SDW
states by considering quantum fluctuations of effective spin moments based on
a formulation by keeping spin rotation symmetry, $\sigma_{z}\rightarrow
\mathbf{n}\cdot \mathbf{\sigma}$ where $\mathbf{n}$ is the SDW order parameter,
$\left \langle \hat{c}_{i}^{\dagger}\mathbf{\sigma}\hat{c}_{i}\right \rangle
=M\mathbf{n}$.

On a honeycomb lattice, after dividing the lattice into two sublattices, $A$
and $B$, the dispersion can be obtained from Eq.(2). In the continuum limit,
the Dirac-like effective Lagrangian describes the low energy fermionic modes
near two points, $\mathbf{k}_{1}=-\frac{2\pi}{3}(1,\frac{1}{\sqrt{3}})$ and
$\mathbf{k}_{2}=\frac{2\pi}{3}(1,\frac{1}{\sqrt{3}}),$ as
\begin{equation}
\mathcal{L}_{f}=\sum_{a}\left[  i\bar{\psi}_{a}\gamma_{\mu}\left(
\partial_{\mu}-iA_{\mu}\right)  \psi_{a}+m_{a}\bar{\psi}_{a}\psi_{a}%
-\delta \Delta_{M}\bar{\psi}_{a}\mathbf{\sigma}\cdot \mathbf{n}\psi_{a}\right]
\label{f'}%
\end{equation}
which describes low energy charged fermionic modes $a=1$ near $\mathbf{k}%
_{1},$
\begin{equation}
\bar{\psi}_{1}=\psi_{1}^{\dagger}\gamma_{0}=(%
\begin{array}
[c]{llll}%
\bar{\psi}_{\uparrow1A}, & \bar{\psi}_{\uparrow1B}, & \bar{\psi}%
_{\downarrow1A}, & \bar{\psi}_{\downarrow1B}%
\end{array}
)
\end{equation}
and $a=2$ near $\mathbf{k}_{2}$,
\begin{equation}
\bar{\psi}_{2}=\psi_{2}^{\dagger}\gamma_{0}=(%
\begin{array}
[c]{llll}%
\bar{\psi}_{\uparrow2B}, & \bar{\psi}_{\uparrow2A}, & \bar{\psi}%
_{\downarrow2B}, & \bar{\psi}_{\downarrow2A}%
\end{array}
).
\end{equation}
The masses of two-flavor fermions are
\begin{equation}
m_{1}=\varepsilon-3\sqrt{3}t^{\prime}%
\end{equation}
and%
\begin{equation}
m_{2}=\varepsilon+3\sqrt{3}t^{\prime}.
\end{equation}
$\gamma_{\mu}$ is defined as $\gamma_{0}=\sigma_{0}\otimes \tau_{z},$
$\gamma_{1}=\sigma_{0}\otimes \tau_{y},$ $\gamma_{2}=\sigma_{0}\otimes \tau_{x}$
with $\sigma_{0}=\left(
\begin{array}
[c]{ll}%
1 & 0\\
0 & 1
\end{array}
\right)  $. $\tau_{x},$ $\tau_{y},$ $\tau_{z}$ are Pauli matrices. $\delta$
$=$ $1$ for $a=1$ and $\delta=-1$ for $a=2$. We have set the Fermi velocity to
be unit $v_{F}=1$.

In\textrm{\ CP}$^{1}$ representation, we may rewrite the effective Lagrangian
of fermions in Eq.(\ref{f'}) as
\begin{equation}
\mathcal{L}_{f}=\sum_{a}\bar{\psi}_{a}^{\prime}\left(  i\gamma_{\mu}%
\partial_{\mu}+\gamma_{\mu}A_{\mu}-\gamma_{\mu}\sigma_{3}a_{\mu}+m_{a}%
-\delta \Delta_{M}\sigma_{3}\right)  \psi_{a}^{\prime} \label{f2}%
\end{equation}
with
\[
\psi_{a}^{^{\prime}}\left(  r,\tau \right)  =U^{\dagger}\left(  r,\tau \right)
\psi_{a}\left(  r,\tau \right)  ,
\]
where $U\left(  r,\tau \right)  $ is a local and time-dependent spin
\textrm{SU(2)} transformation defined by
\[
U^{\dagger}\left(  r,\tau \right)  \mathbf{n\cdot \sigma}U\left(  r,\tau \right)
=\sigma_{3}.
\]
And $a_{\mu}$\ is introduced as an assistant gauge field as
\[
i\sigma_{3}a_{\mu}\equiv U^{\dagger}\left(  r,\tau \right)  \partial_{\mu
}U\left(  r,\tau \right)  .
\]

An important property of above model in Eq.(\ref{f2}) is the current anomaly.
The vacuum expectation value of the fermionic current
\begin{equation}
J_{a,\sigma}^{\mu}=i\langle \bar{\psi}_{a,\sigma}\gamma^{\mu}\psi_{a,\sigma
}\rangle
\end{equation}
can be defined by
\begin{equation}
J_{a,\sigma}^{\mu}=i\{ \gamma^{\mu}[(i\hat{D}+im_{a,\sigma}\mathbf{)}%
^{\dagger}(i\hat{D}+im_{a,\sigma}\mathbf{)}]^{-1}(i\hat{D}+im_{a,\sigma
}\mathbf{)}^{\dagger}\}
\end{equation}
where
\begin{equation}
\hat{D}=\gamma_{\mu}(\partial_{\mu}-iA_{\mu}+i\sigma a_{\mu})
\end{equation}
and the mass terms are $m_{a,\sigma}=m_{a}-\delta \Delta_{M}\sigma$. The
topological current $J_{a,\sigma}^{\mu}$ is obtained to be
\begin{equation}
J_{a,\sigma}^{\mu}=\frac{1}{2}\frac{1}{4\pi}\frac{m_{a,\sigma}}{|m_{a,\sigma
}|}\epsilon^{\mu \nu \lambda}(\partial_{\nu}A_{\lambda}-\sigma \partial_{\nu
}a_{\lambda}).
\end{equation}
Then we derive the CSH terms as\cite{redlich,cs}
\begin{equation}
\mathcal{L}_{CSH}=-i\sum_{a,\sigma}(A_{\mu}-\sigma a_{\mu})J_{a,\sigma}^{\mu}.
\end{equation}

To make an explicit description of SDWs, we introduce the $\mathcal{K}$-matrix
formulation that has been used to characterize FQH fluids
successfully\cite{Kmat}. Now the CSH term is written as
\begin{equation}
\mathcal{L}_{CSH}=-i\sum_{I,J}\frac{\mathcal{K}_{IJ}}{4\pi}\varepsilon^{\mu
\nu \lambda}a_{\mu}^{I}\partial_{\nu}a_{\lambda}^{J} \label{cs1}%
\end{equation}
where $\mathcal{K}$ is 2-by-2 matrix, $a_{\mu}^{I=1}=A_{\mu}$ and $a_{\mu
}^{I=2}=a_{\mu}.$ The "charge" of $A_{\mu}$ and $a_{\mu}$ are defined by $q$
and $q_{s}$, respectively.

Thus for different SDW orders with the same order parameter $M$, we have
different $\mathcal{K}$-matrices : for $m_{1},$ $m_{2}>\Delta_{M},$
\begin{equation}
\mathcal{K}=\left(
\begin{array}
[c]{ll}%
2 & 0\\
0 & 2
\end{array}
\right)  ;
\end{equation}
for $m_{2}>\Delta_{M}>m_{1}$,
\begin{equation}
\mathcal{K}=\left(
\begin{array}
[c]{ll}%
1 & 1\\
1 & 1
\end{array}
\right)  ;
\end{equation}
for $m_{1},$ $m_{2}<\Delta_{M}$,
\begin{equation}
\mathcal{K}=0.
\end{equation}
This results are consistent to those in Ref.[1].

\end{document}